\RequirePackage{lineno}
\documentclass[10pt]{IEEEtran} %technote
\usepackage{epsfig,amsmath,amsfonts,cite}
\usepackage{threeparttable}
\usepackage{lipsum}
\usepackage{color}
\usepackage{graphicx}
\usepackage{lipsum}
\usepackage{algorithm,algorithmic}
\usepackage{epsfig,amsmath,amsfonts,cite}
\usepackage{amsthm}
\usepackage{booktabs}
\newcommand{\thickhline}{\noalign{\hrule height 1.0pt}}
\DeclareMathAlphabet\mathbfcal{OMS}{cmsy}{b}{n}

\usepackage{multirow}%,multicol
\usepackage{multicol}
\newtheorem{example}{Example}%[section]
\newtheorem{definition}{Definition}%[section]
\ifCLASSINFOpdf
 
\else
\fi

\begin{document}

\title{Enabling High-Dimensional Hierarchical Uncertainty Quantification by ANOVA and Tensor-Train Decomposition}

\author{Zheng Zhang, Xiu Yang, Ivan V. Oseledets, George Em Karniadakis, and~Luca~Daniel\\
Accepted by IEEE Trans. Computer-Aided Design of Integrated Circuits and Systems

\thanks{Some preliminary results of this work have been reported in~\cite{zzhang_cicc2014}. This work was funded by the MIT-SkolTech program. I. Oseledets was also supported by the Russian Science Foundation under Grant 14-11-00659.}% Cooperative Agreement between the Masdar Institute of Science and Technology, Abu Dhabi, UAE and the Massachusetts Institute of Technology (MIT), Cambridge, MA, USA (Reference No. 196F/002/707/102f/70/9374).}      
\thanks{Z. Zhang and L. Daniel are with the Research Laboratory of Electronics, Massachusetts Institute of Technology (MIT), Cambridge, MA 02139, USA (e-mail: z\_zhang@mit.edu, luca@mit.edu).}
\thanks{X. Yang was with the Division of Applied Mathematics, Brown University, Providence, RI 02912. Now he is with the Pacific Northwest National Laboratory, Richland, WA 99352, USA (e-mail: xiu.yang@pnnl.gov).}
\thanks{G. Karniadakis is with the Division of Applied Mathematics, Brown University, Providence, RI 02912, USA (e-mail: george\_karniadakis@brown.edu).}
\thanks{Ivan V. Oseledets is with the Skolkovo Institute of Science and Technology, Skolkovo 143025, Russia (e-mail: ivan.oseledets@gmail.com).}
%\thanks{Copyright (c) 2015 IEEE. Personal use of this material is permitted. However, permission to use this material for any other purposes must be obtained from the IEEE by sending an email to pubs-permissions@ieee.org.}
}

\markboth{IEEE TRANSACTIONS ON COMPUTER-AIDED DESIGN OF INTEGRATED CIRCUITS AND SYSTEMS, ~Vol. ~XX, No.~XX,~XX~2015}{ZHANG \MakeLowercase{\textit{et al.}}: Enabling Hierarchical UQ by ANOVA and TT Decomposition}

\maketitle

\begin{abstract}
Hierarchical uncertainty quantification can reduce the computational cost of stochastic circuit simulation by employing spectral methods at different levels. This paper presents an efficient framework to simulate hierarchically some challenging stochastic circuits/systems that include high-dimensional subsystems. Due to the high parameter dimensionality, it is challenging to both extract surrogate models at the low level of the design hierarchy and to handle them in the high-level simulation. In this paper, we develop an efficient ANOVA-based stochastic circuit/MEMS simulator to extract efficiently the surrogate models at the low level. In order to avoid the curse of dimensionality, we employ tensor-train decomposition at the high level to construct the basis functions and Gauss quadrature points. As a demonstration, we verify our algorithm on a stochastic oscillator with four MEMS capacitors and $184$ random parameters. This challenging example is simulated efficiently by our simulator at the cost of only $10$ minutes in MATLAB on a regular personal computer. 
\end{abstract}


\IEEEpeerreviewmaketitle

\section{Introduction}

\IEEEPARstart{P}{rocess} variations have become a major concern in submicron and nano-scale chip design~\cite{variation2008,LiYu:2014dac,LiYu:2014date,LiYu:2013date,LiYu:2012ISQED}. In order to improve chip performances, it is highly desirable to develop efficient stochastic simulators to quantify the uncertainties of integrated circuits and microelectromechanical systems (MEMS). Recently, stochastic spectral methods~\cite{sfem,gPC2002,col:2005,Ivo:2007,Nobile:2008,Nobile:2008_2} have emerged as a promising alternative to Monte Carlo techniques~\cite{SingheeR10}. The key idea is to represent the stochastic solution as a linear combination of some basis functions (e.g., generalized polynomial chaos~\cite{gPC2002}), and then compute the solution by stochastic Galerkin~\cite{sfem}, stochastic collocation~\cite{col:2005,Ivo:2007,Nobile:2008,Nobile:2008_2} or stochastic testing~\cite{zzhang:tcad2013,zzhang:tcas2_2013,zzhang:iccad_2013} methods. Due to the fast convergence rate, such techniques have been successfully applied in the stochastic analysis of integrated circuits~\cite{zzhang:tcad2013,zzhang:tcas2_2013,zzhang:iccad_2013,Strunz:2008,Tao:2007,Manfredi:tcas1_2014,Pulch:2011_1}, VLSI interconnects~\cite{Stievano:2011,cmpt2012,Vrudhula:2006,Tarek_ISQED:11,Tarek_DAC:08}, electromagnetic~\cite{sMOR2012} and MEMS devices~\cite{MEMS_uq_jmems09,zzhang_cicc2014}, achieving significant speedup over Monte Carlo when the parameter dimensionality is small or medium.

Since many electronic systems are designed in a hierarchical way, it is possible to exploit such structure and simulate a complex circuit by hierarchical uncertainty quantification~\cite{zzhang:tcad2014}\footnote{Design hierarchy can be found in many engineering fields. In the recent work~\cite{Ng:2014} a hierarchical stochastic analysis and optimization framework based on multi-fidelity models~\cite{Ng:2014_opt,Allaire:2014} was proposed for aircraft design.}. Specifically, one can first utilize stochastic spectral methods to extract surrogate models for each block. Then, circuit equations describing the interconnection of blocks may be solved with stochastic spectral methods by treating each block as a single random parameter. Typical application examples include (but are not limited to) analog/mixed-signal systems (e.g., phase-lock loops) and MEMS/IC co-design. In our preliminary conference paper~\cite{zzhang_cicc2014}, this method was employed to simulate a low-dimensional stochastic oscillator with $9$ random parameters, achieving $250\times$ speedup over the hierarchical Monte-Carlo method proposed in~\cite{Felt:1996}.

\textbf{Paper Contributions.} This paper extends the recently developed hierarchical uncertainty quantification method~\cite{zzhang:tcad2014} to the challenging cases that include subsystems with high dimensionality (i.e., with a large number of parameters). Due to such high dimensionality, it is too expensive to extract a surrogate model for each subsystem by any standard stochastic spectral method. It is also non-trivial to perform high-level simulation with a stochastic spectral method, due to the high-dimensional integration involved when computing the basis functions and Gauss quadrature rules for each subsystem. In order to reduce the computational cost, this work develops some fast numerical algorithms to accelerate the simulations at both levels:
\begin{itemize}
	\item At the low level, we develop a sparse stochastic testing simulator based on adaptive anchored ANOVA~\cite{anchor_ANOVA_xiu:2012,HDMR:1999,anchor_ANOVA_Griebel:2010,ANOVA_zqzhang:2012,anchor_ANOVA_xma:2010} to efficiently simulate each subsystem. This approach exploits the sparsity on-the-fly, and it turns out to be suitable for many circuit and MEMS problems. This algorithm was reported in our preliminary conference paper~\cite{zzhang_cicc2014} and was used for the global sensitivity analysis of analog integrated circuits.
	\item In the high-level stochastic simulation, we accelerate the three-term recurrence relation~\cite{Walter:1982} by tensor-train decomposition~\cite{Ivan:tt_2011,Ivan:tt_svd, Ivan:tt_across}. Our algorithm has a linear complexity with respect to the parameter dimensionality, generating a set of basis functions and Gauss quadrature points with high accuracy (close to the machine precision). This algorithm was not reported in~\cite{zzhang_cicc2014}.	
\end{itemize}

\section{Background Review}
This section first reviews the recently developed stochastic testing circuit/MEMS simulator~\cite{zzhang:tcad2013,zzhang:tcas2_2013,zzhang:iccad_2013} and hierarchical uncertainty quantification~\cite{zzhang:tcad2014}. Then we introduce some background about tensor and tensor decomposition.

\subsection{Stochastic Testing Circuit/MEMS Simulator}
 Given a circuit netlist (or a MEMS 3D schematic file), device models and process variation descriptions, one can set up a stochastic differential algebraic equation:
\begin{equation}
\label{eq:sdae}
\begin{array}{l}
 \displaystyle{\frac{{d\vec q\left( {\vec x( {t,\vec \xi } ),\vec \xi } \right)}}{{dt}} }+ \vec f\left( {\vec x( {t,\vec \xi } ),\vec \xi }, u(t) \right) = 0 
 \end{array}
\end{equation}
where $\vec u(t)$ is the input signal, ${\vec \xi}$=$[\xi_1,\cdots,\xi_d]\in\Omega \subseteq\mathbb{R}^d$ are $d$ mutually independent random variables describing process variations. The joint probability density function of $\vec \xi$ is
\begin{equation}
\label{PDF}
\rho(\vec \xi)=\prod\limits_{k = 1}^d {\rho _{k } \left( \xi_k \right)},
\end{equation}
where ${\rho _{k } \left( \xi_k \right)}$ is the marginal probability density function of $\xi_k \in \Omega_k$. In circuit analysis, ${\vec x}$$\in $$\mathbb{R}^n$ denotes nodal voltages and branch currents; ${\vec q}$$\in$$ \mathbb{R}^n$ and ${\vec f}$$\in $$\mathbb{R}^n$ represent charge/flux and current/voltage, respectively. In MEMS analysis, Eq. (\ref{eq:sdae}) is the equivalent form of a commonly used $2$nd-order differential equation~\cite{zzhang_cicc2014,zzhang:JMEMS2014}; ${\vec x}$ includes displacements, rotations and their first-order derivatives with respect to time $t$. 
%The joint probability density function of $\vec \xi$ is
%\begin{equation}
%\label{PDF}
%\rho(\vec \xi)=\prod\limits_{k = 1}^d {\rho _{k } \left( \xi_k \right)},
%\end{equation}
%where ${\rho _{k } \left( \xi_k \right)}$ is the marginal density of $\xi_k \in \Omega_k \subseteq \mathbb{R}$. 

When $\vec x({\vec \xi},t)$ has a bounded variance and smoothly depends on $\vec \xi$, we can approximate it by a truncated generalized polynomial chaos expansion~\cite{gPC2002}
\begin{equation}	
\label{gpcExpan}
\vec x(t,\vec \xi ) \approx \tilde x(t,\vec \xi) =\sum\limits_{\vec \alpha \in {\cal P}} {\hat x_{\vec \alpha} (t)H_{\vec \alpha}(\vec \xi )} 
\end{equation}
where $\hat x_{\vec \alpha} (t)\in \mathbb{R}^n$ denotes a coefficient indexed by vector ${\vec \alpha}=[\alpha_1,\cdots,\alpha_d]\in \mathbb{N}^d$, and the basis function $H_{\vec \alpha}(\vec \xi )$ is a multivariate polynomial with the highest order of $\xi_i$ being $\alpha_i$. In practical implementations, it is popular to set the highest total degree of the polynomials as $p$, then ${\cal P}=\{\vec \alpha |\; \alpha_k\in \mathbb{N},\; 0\leq {\alpha _1}+\cdots+\alpha_d \leq p\}$ and the total number of basis functions is
\begin{equation}
\label{Kvalue}
K = \left( \begin{array}{l}
 p + d \\ 
 \;\;p \\ 
 \end{array} \right) = \frac{{(p + d)!}}{{p!d!}}.
\end{equation}
For any integer $j$ in $[1,K]$, there is a one-to-one correspondence between $j$ and $\vec \alpha$. As a result, we can denote a basis function as $H_j(\vec \xi)$ and rewrite (\ref{gpcExpan}) as
\begin{equation}	
\label{gpcExpan_k}
\vec x(t,\vec \xi ) \approx \tilde x(t,\vec \xi) =\sum\limits_{j=1}^K {\hat x_{j} (t)H_{j}(\vec \xi )} .
\end{equation}
%Since all components of $\vec \xi$ are mutually independent, one can construct a set of univariate orthonormal polynomials for each $\xi_i$ and then obtain $H_{\vec \alpha}(\vec \xi )$, as detailed in Section II-A of~\cite{zzhang:iccad_2013}.

In order to compute $\vec x(t,\vec \xi )$, stochastic testing~\cite{zzhang:tcad2013,zzhang:tcas2_2013,zzhang:iccad_2013} substitutes $\tilde x(t,\vec \xi)$ into (\ref{eq:sdae}) and forces the residual to zero at $K$ testing samples of $\vec \xi$. This gives a deterministic differential algebraic equation of size $nK$
\begin{align}
\label{ST:forced}
\frac{{d\textbf{q}(\hat{\textbf{x}}(t))}}{{dt}} + \textbf{f}\left(\hat{\textbf{x}}(t),u(t)\right) =0,
\end{align}
where the state vector $\hat{\textbf{x}}(t)$ contains all coefficients in (\ref{gpcExpan}). Stochastic testing then solves Eq. (\ref{ST:forced}) with a linear complexity of $K$ and with adaptive time stepping, and it has shown higher efficiency than standard stochastic Galerkin and stochastic collocation methods in circuit simulation~\cite{zzhang:tcad2013,zzhang:tcas2_2013}.

\subsubsection{Constructing Basis Functions}
The basis function $H_{\vec \alpha}(\vec \xi )$ is constructed as follows (see Section II of~\cite{zzhang:iccad_2013} for details):
\begin{itemize}
	\item First, for $\xi_i$ one constructs a set of degree-$\alpha_i$ orthonormal univariate polynomials $\left\{ \varphi_{\alpha_i}^i(\xi_i)\right\}_{\alpha_i=0}^p$ according to its marginal probability density $\rho_i(\xi_i)$.
	\item Next, based on the obtained univariate polynomials of each random parameter one constructs the multivariate basis function:  $H_{\vec \alpha}(\vec \xi )$$=$$\prod_{i=1}^d{\varphi_{\alpha_i}^i(\xi_i)}$.
\end{itemize}

The obtained basis functions are orthonormal polynomials in the multi-dimensional parameter space $\Omega$ with the density measure $\rho(\vec \xi)$. As a result, some statistical information can be easily obtained. For example, the mean value and variance of $\vec {x}(t,\vec \xi)$ are $\hat x_{0}(t)$ and $\sum\limits_{\vec{\alpha} \neq 0} \left(\hat {x}_{\vec \alpha}(t)\right)^2$, respectively.

\subsubsection{Testing Point Selection}
The selection of testing points influence the numerical accuracy of the simulator. In stochastic testing, the testing points $\{ \vec \xi ^j  \}_{j=1}^K$ are selected by the following two steps (see Section III-C of \cite{zzhang:tcad2013} for details):
\begin{itemize}
	\item First, compute a set of multi-dimensional quadrature points. Such quadrature points should give accurate results for evaluating the numerical integration of any multivariate polynomial of $\vec \xi$ over $\Omega$ [with density measure $\rho(\vec \xi)$] when the polynomial degree is $\leq 2p$.
	\item Next, among the obtained quadrature points, we select the $K$ samples with the largest quadrature weights under the constraint that $\mathbf{V}\in \mathbb{R}^{K\times K}$ is well-conditioned. The $(j,k)$ element of $\mathbf{V}$ is $H_k(\vec \xi^j)$.
\end{itemize}

%Stochastic testing can be hundreds of times faster than Monte Carlo when $d$ is small or medium. However, for high-dimensional prolbems, lots of basis functions and samples are required, leading to curse of dimensionality. This is a common problem of all stochastic spectral methods~\cite{col:2005,sfem,gPC2002}.

\subsection{Hierarchical Uncertainty Quantification}
\label{subsec:huq}
\begin{figure}[t]
	\centering
		\includegraphics[width=3.3in]{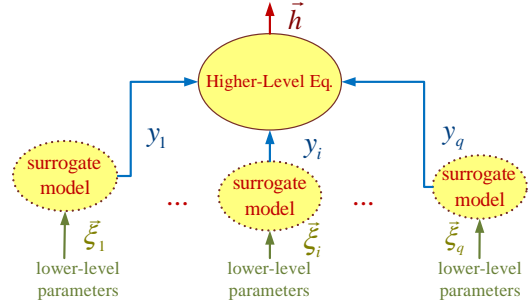} 
\caption{Demonstration of hierarchical uncertainty quantification.}
	\label{fig:hierarchical}
\end{figure}
Consider Fig.~\ref{fig:hierarchical}, where an electronic system has $q$ subsystems. The output $y_i$ of a subsystem is influenced by some process variations $\vec \xi_i\in \mathbb{R}^{d_i}$, and the output $\vec h$ of the whole system depends on all random parameters $\vec \xi_i$'s. For simplicity, in this paper we assume that $y_i$ only depends on $\vec \xi_i$ and does not change with time or frequency. Directly simulating the whole system can be expensive due to the large problem size and high parameter dimensionality. If $y_i$'s are mutually independent and smoothly dependent on $\vec \xi_i$'s, we can accelerate the simulation in a hierarchical way~\cite{zzhang:tcad2014}:
\begin{itemize}
	\item First, perform low-level uncertainty quantification. We use stochastic testing to simulate each block, obtaining a generalized polynomial expansion for each $y_i$. In this step we can also employ other stochastic spectral methods such as stochastic Galerkin or stochastic collocation.
	\item Next, perform high-Level uncertainty quantification. By treating $y_i$'s as the inputs of the high-level equation, we use stochastic testing again to efficiently compute $\vec h$. Since $y_i$ has been assumed independent of time and frequency, we can treat it as a random parameter. %, and thus it is a generalized polynomial chaos expansion of $\vec \xi_i$.
\end{itemize}

In order to apply stochastic spectral methods at the high level, we need to compute a set of \textit{specialized} orthonormal polynomials and Gauss quadrature points/weights for each input random parameter. For the sake of numerical stability, we define a zero-mean unit-variance random variable $\zeta_i$ for each subsystem, by shifting and scaling $y_i$. The intermediate variables $\vec \zeta=[\zeta_1, \cdots, \zeta_q]$ are used as the random parameters in the high-level equation. Dropping the subscript for simplicity, we denote a general intermediate-level random parameter by $\zeta$ and its probability density function by $\rho (\zeta)$ (which is actually unknown), then we can construct $p+1$ orthogonal polynomials $\{\pi_j(\zeta)\}_{j=0}^p$ via a three-term recurrence relation~\cite{Walter:1982} 
\begin{equation}
\label{recurrence}
\begin{array}{l}
 \pi _{j + 1} (\zeta) = \left( {\zeta - \gamma _j } \right)\pi _j (\zeta) - \kappa _j \pi _{j - 1} (\zeta), \\ 
 \pi _{- 1} (\zeta) = 0,\;\;\pi _0 (\zeta) = 1,\;\;j = 0, \cdots , p-1  
 \end{array} 
\end{equation}
with 
\begin{equation}
\label{int_cal}
\begin{array}{l}
 \displaystyle{\gamma _j  = \frac{{\int\limits_{\mathbb{R}} {\zeta\pi _j^2 (\zeta)\rho (\zeta)d\zeta} }}{{\int\limits_{\mathbb{R}} {\pi _j^2 (\zeta)\rho (\zeta)d\zeta} }}}, \displaystyle{\;\kappa _{j+1}  = \frac{{\int\limits_{\mathbb{R}} {\pi _{j+1}^2 (\zeta)\rho (\zeta)d\zeta} }}{{\int\limits_{\mathbb{R}} {\pi _{j}^2 (\zeta)\rho (\zeta)d\zeta} }}}
 \end{array}
\end{equation}
and $\kappa_0=1$, where $\pi_j(\zeta)$ is a degree-$j$ polynomial with a leading coefficient 1. The first $p+1$ univariate basis functions can be obtained by normalization:
\begin{equation}
\phi _j (\zeta) = \frac{{\pi _j (\zeta)}}{{\sqrt {\kappa _0 \kappa _1  \cdots \kappa _j } }}, \; {\rm for}\; j=0,1,\cdots, p.
\end{equation} 
The parameters $\kappa_j$'s and $\gamma_j$'s can be further used to form a symmetric tridiagonal matrix $\mathbf{J} \in \mathbb{R}^{(p+1)\times (p+1)}$: 
\begin{equation}
\label{eq:jmatrix}
\mathbf{J}\left( {j,k} \right) = \left\{ \begin{array}{l}
 \gamma _{j - 1} ,\;{\rm{if}}\;j = k \\ 
 \sqrt {\kappa _j } ,\;{\rm{if}}\;k = j + 1 \\ 
 \sqrt {\kappa _k } ,\;{\rm{if}}\;k = j - 1 \\ 
 0,\;{\rm{otherwise}} \\ 
 \end{array} \right.\;{\rm{for}}\;1 \le j,k \le p + 1.
\end{equation}
Let $\mathbf{J} = \mathbf{U}\Sigma \mathbf{U}^T$ be an eigenvalue decomposition, where $\mathbf{U}$ is a unitary matrix. The $j$-th quadrature point and weight of $\zeta$ are $\Sigma(j,j)$ and $\left(\mathbf{U}(1,j)\right)^2$, respectively~\cite{Golub:1969}. %With the orthonormal polynomials and Gauss quadrature rule for each $\zeta_i$, we can compute $\vec h$ by stochastic testing~\cite{zzhang:tcad2013,zzhang:tcas2_2013,zzhang:iccad_2013}.

\textit{Challenges in High Dimension.} When $d_i$ is large, it is difficult to implement hierarchical uncertainty quantification. First, it is non-trivial to obtain a generalized polynomial chaos expansion for $y_i$, since a huge number of basis functions and samples are required to obtain $y_i$. Second, when high accuracy is required, it is expensive to implement (\ref{recurrence}) due to the non-trivial integrals when computing $\kappa_j$ and $\gamma_j$. Since the density function of $\zeta_i$ is unknown, the integrals must be evaluated in the domain of $\vec \xi_i$, with a cost growing exponentially with $d_i$ when a deterministic quadrature rule is used.

\subsection{Tensor and Tensor Decomposition}
%As a generalization of matrix, tensor has emerged as an efficient tool for high-dimensional problems in many engineering fields~\cite{tensor:Kolda2008,tensor:Vasilescu2002,tensor:latentvar,tensor:gelerkin,qtt:sc}. 
\begin{definition}[\textbf{Tensor}] 
A tensor $\mathbfcal{A}\in \mathbb{R}^{N_1\times N_2\times \cdots \times N_d} $ is a multi-mode (or multi-way) data array. The mode (or way) is $d$, the number of dimensions. The size of the $k$-th dimension is $N_k$. An element of the tensor is $\mathbfcal{A}(i_1,\cdots, i_d)$, where the positive integer $i_k$ is the index for the $k$-th dimension and $1\leq i_k\leq N_k$. The total number of elements of $\mathbfcal{A}$ is $N_1\times \cdots \times N_d$.
\end{definition}

\begin{figure}[t]
	\centering
		\includegraphics[width=3.3in]{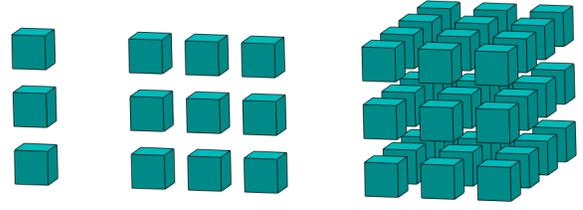} 
\caption{Demonstration of a vector (left), a matrix (center) and a $3$-mode tensor (right).}
	\label{fig:tensor}
\end{figure}

As a demonstration, we have shown a vector ($1$-mode tensor) in $\mathbb{R}^{3\times 1}$, a matrix ($2$-mode tensor) in $\mathbb{R}^{3\times 3}$ and a $3$-mode tensor in $\mathbb{R}^{3\times 3\times 4}$ in Fig.~\ref{fig:tensor}, where each small cube represents a scalar.

\begin{definition}[\textbf{Inner Product of Two Tensors}] 
For $\mathbfcal{A}, \mathbfcal{B}\in \mathbb{R}^{N_1\times N_2\times \cdots \times N_d} $, their inner product is defined as the sum of their element-wise product
\begin{equation}
\left\langle {\mathbfcal{A},\mathbfcal{B}} \right\rangle  =  \sum\limits_{i_1,\cdots,i_d } {\mathbfcal{A}\left( {i_1 , \cdots i_d } \right)\mathbfcal{B}\left( {i_1 , \cdots i_d } \right)}. 
\end{equation}
\end{definition}

\begin{definition}[\textbf{Frobenius Norm of A Tensor}] 
For $\mathbfcal{A}\in \mathbb{R}^{N_1\times N_2\times \cdots \times N_d} $, its Frobenius norm is defined as
\begin{equation}
\left\|\mathbfcal{A} \right\|_F= \sqrt{\left\langle {\mathbfcal{A},\mathbfcal{A}} \right\rangle }. 
\end{equation}
\end{definition}

%\begin{definition}[\textbf{Outer Product}] 
%Given vectors $\mathbf{v}^{(1)} \in \mathbb{R}^{N_1}$, $\mathbf{v}^{(2)} \in \mathbb{R}^{N_2}$, $\cdots$, $\mathbf{v}^{(d)} \in \mathbb{R}^{N_d}$, let $\mathbfcal{C}=\mathbf{v}^{(1)}  \circ \mathbf{v}^{(2)}  \cdots  \circ \mathbf{v}^{(1)}$ denote their outer product, then  
%\begin{equation}
%\label{outer}
%
%\end{equation}
%where $\mathbf{v}^{(k)}(i_k)$ denotes the $k$-th element of vector $\mathbf{v}^{(k)}$. $\mathbfcal{C}(i_1,\cdots, i_d)= \prod_{k=1}^{d}\mathbf{v}^{(k)}(i_k), $
%\end{definition}

\begin{definition}[\textbf{Rank-One Tensors}] 
A $d$-mode tensor $\mathbfcal{A}\in \mathbb{R}^{N_1 \times \cdots \times N_d}$ is rank one if it can be written as the outer product of $d$ vectors
\begin{equation}
\label{outer}
\mathbfcal{A}=\mathbf{v}^{(1)}  \circ \mathbf{v}^{(2)}  \cdots  \circ \mathbf{v}^{(d)} ,  \; {\rm with} \; \mathbf{v}^{(k)}\in \mathbb{R}^{N_k}
\end{equation}
where $\circ$ denotes the outer product operation. This means that 
\begin{equation}
\mathbfcal{A}(i_1,\cdots, i_d)= \prod_{k=1}^{d}\mathbf{v}^{(k)}(i_k)\; {\rm for} \; {\rm all}\; 1\leq i_k \leq N_k.
\end{equation}
Here $\mathbf{v}^{(k)}(i_k)$ denotes the $i_k$-th element of vector $\mathbf{v}^{(k)}$.
\end{definition}

\begin{definition}[\textbf{Tensor Rank}] 
The rank of $\mathbfcal{A}\in \mathbb{R}^{N_1 \times \cdots \times N_d}$ is the smallest positive integer $\bar r$, such that 
\begin{equation}
\label{cp}
\mathbfcal{A}=\sum_{j=1}^{\bar r}\mathbf{v}_j^{(1)}  \circ \mathbf{v}_j^{(2)}  \cdots  \circ \mathbf{v}_j^{(d)} ,  \; {\rm with} \; \mathbf{v}_j^{(k)}\in \mathbb{R}^{N_k}.
\end{equation}
\end{definition}

It is attractive to perform tensor decomposition: given a small integer $r<\bar r$, approximate $\mathbfcal{A}\in \mathbb{R}^{N_1 \times \cdots \times N_d}$ by a rank-$r$ tensor. Popular tensor decomposition algorithms include canonical decomposition~\cite{cp:Hitchcock,cp:Carroll,cp:Kiers} and Tuker decomposition~\cite{tucker:1966, tucker:2000}. Canonical tensor decomposition aims to approximate $\mathbfcal{A}$ by the sum of $r$ rank-$1$ tensors [in the form of (\ref{cp})] while minimizing the approximation error, which is normally implemented with alternating least square~\cite{cp:Carroll}. This decomposition scales well with the dimensionality $d$, but it is ill-posed for $d\geq 3$~\cite{tensor_ill:lim}. Tucker decomposition aims to represent a tensor by a small core tensor and some matrix factors~\cite{tucker:1966, tucker:2000}. This decomposition is based on singular value decomposition. It is robust, but the number of elements in the core tensor still grows exponentially with $d$.

Alternatively, tensor-train decomposition~\cite{Ivan:tt_svd,Ivan:tt_2011,Ivan:tt_across} approximates $\mathbfcal{A}\in \mathbb{R}^{N_1 \times \cdots \times N_d}$ by a low-rank tensor $\hat{\mathbfcal{A}}$ with
\begin{equation}
\label{eq:tt_decomp_def}
\hat {\mathbfcal{A}}\left( {i_1 , \cdots i_d } \right) = \mathbfcal{G}_1 \left( {:,i_1 ,:} \right)\mathbfcal{G}_2 \left( {:,i_1 ,:} \right) \cdots \mathbfcal{G}_d \left( {:,i_d ,:} \right).
\end{equation}
Here $\mathbfcal{G}_k\in \mathbb{R}^{r_{k-1}\times N_k \times r_k}$, and $r_0=r_d=1$. By fixing the second index $i_k$, $\mathbfcal{G}_k(:,i_k,:)$$\in $$\mathbb{R}^{r_{k-1}\times  r_k}$ becomes a matrix (or vector when $k$ equals $1$ or $d$). To some extent, tensor-train decomposition have the advantages of both canonical tensor decomposition and Tuker decomposition: it is robust since each core tensor is obtained by a well-posed low-rank matrix decomposition~\cite{Ivan:tt_svd,Ivan:tt_2011,Ivan:tt_across}; it scales linearly with $d$ since storing all core tensors requires only $O(Nr^2d)$ memory if we assume $N_k=N$ and $r_k=r$ for $k=1,\cdots, d-1$. Given an error bound $\epsilon$, the tensor train decomposition in (\ref{eq:tt_decomp_def}) ensures
\begin{equation}
\left\| {\mathbfcal{A} - \hat {\mathbfcal{A}}} \right\|_F  \le \varepsilon \left\| {\mathbfcal{A} } \right\|_F
\end{equation}
while keeping $r_k$'s as small as possible~\cite{Ivan:tt_2011}.

\begin{definition}[\textbf{TT-Rank}] 
In tensor-train decomposition (\ref{eq:tt_decomp_def}) $\mathbfcal{G}_k\in \mathbb{R}^{r_{k-1}\times N_k \times r_k}$ for $k=1,\cdots d$. The vector $\vec r=[r_0,r_1,\cdots, r_d]$ is called TT-rank.
\end{definition}

Recently, tensor decomposition has shown promising applications in high-dimensional data and image compression~\cite{tensor:bigdata,tensor:jmsun,tensor:Kolda2008,tensor:Vasilescu2002}, and in machine learning~\cite{tensor:suplearning,tensor:latentvar}. In the uncertainty quantification community, some efficient high-dimensional stochastic PDE solvers have been developed based on canonical tensor decomposition~\cite{doostan:2009,Nouy:2010,Nouy:2009} (which is called ``Proper Generalized Decomposition" in some papers) and tensor-train decomposition~\cite{tensor:gelerkin,Schwab:2014,qtt:sc,Dolgov:2014}. In~\cite{Marzouk:2014}, a spectral tensor-train decomposition is proposed for high-dimensional function approximation.

\section{ANOVA-Based Surrogate Model Extraction}
\label{sec:ANOVA}
In order to accelerate the low-level simulation, this section develops a sparse stochastic circuit/MEMS simulator based on anchored ANOVA (analysis of variance). Without of loss of generality, let  $y=g(\vec \xi)$ denote the output of a subsystem. We assume that $y$ is a smooth function of the random parameters $\vec \xi \in \Omega \subseteq \mathbb{R}^d$ that describe the process variations.

\subsection{ANOVA and Anchored ANOVA Decomposition}
\subsubsection{ANOVA} With ANOVA decomposition~\cite{ANOVA_sobol:2001,HDMR:1999}, $y$ can be written as
\begin{equation}
\label{eq:ANOVA}
y=g(\vec \xi ) = \sum\limits_{{\it s} \subseteq {\cal I}} {g_{\it s} (\vec \xi _{\it s} )},
\end{equation}
where ${\it s}$ is a subset of the full index set ${\cal I}=\left\{ 1, 2,\cdots, d\right\}$. Let $\bar{\it s}$ be the complementary set of $s$ such that ${\it s}\cup \bar{\it s}={\cal I}$ and ${\it s}\cap \bar{\it s}=\emptyset$, and let $|{\it s}|$ be the number of elements in ${\it s}$. When ${\it s}=\left \{i_1,\cdots, i_{|{\it s}|} \right\}\neq \emptyset$, we set $\Omega_{\it s}=\Omega_{i_1}\otimes\cdots \otimes \Omega_{i_{|{\it s}|}}$, $\vec \xi _{\it s}=[\xi_{i_1},\cdots,\xi_{i_{|{\it s}|}}]\in \Omega_{\it s}$ and have the Lebesgue measure
\begin{equation}
\label{eq:Lebesgure}
d\mu ( {\vec \xi _{\bar s} } ) = \prod\limits_{k \in \bar s} {\left( {\rho _k \left( {\xi _k } \right)d\xi _k } \right)}.
\end{equation} 
Then, $g_{\it s} (\vec \xi _{\it s} )$ in ANOVA decomposition (\ref{eq:ANOVA}) is defined recursively by the following formula
\begin{equation}
\label{eq:ANOVA_term}
g_{\it s} (\vec \xi _{\it s} ) = \left\{ \begin{array}{l}
 \mathbb{E}\left( {g( {\vec \xi } )} \right) = \int\limits_\Omega  {g( {\vec \xi })d\mu ( {\vec \xi } )}  = g_0 ,\;{\rm{if}}\;{\it s} = \emptyset  \\ 
 \hat g_{\it s} (\vec \xi _{\it s} ) - \sum\limits_{{\it t} \subset {\it s}} {g_{\it t} ( {\vec \xi _{\it t} } )\;} ,\;\;{\rm{if}}\;{\it s} \ne \emptyset. 
 \end{array} \right.
\end{equation}
Here $\mathbb{E}$ is the expectation operator, $\hat g_{\it s} (\vec \xi _{\it s} ) = \int\limits_{\Omega _{\bar {\it s}} } {g( {\vec \xi } )d\mu ( {\vec \xi _{\bar {\it s}} } )} $, and the integration is computed for all elements except those in $\vec \xi_{\it s}$. From (\ref{eq:ANOVA_term}), we have the following intuitive results:
\begin{itemize}
	\item $g_0$ is a constant term;
	\item if ${\it s}$$=$$\{j \}$, then $\hat g_{\it s} (\vec \xi _{\it s} )=\hat g_{\{j\}} (\xi _{j} )$, $g_{\it s} (\vec \xi _{\it s} )=g_{\{j\}} (\xi _{j} )$ $=$ $\hat g_{\{j\}} (\xi _{j} )-g_0$; 
	\item if ${\it s}$$=$$\{j,k\}$ and $j<k$, then $\hat g_{\it s} (\vec \xi _{\it s} )=\hat g_{\{j,k\}} (\xi_j,\xi_k)$ and $g_{\it s} (\vec \xi _{\it s} )=\hat g_{\{j,k\}} (\xi_j,\xi_k)- g_{\{j\}}(\xi_j)-g_{\{k\}}(\xi_k)-g_0$;
	\item both $\hat g_{\it s} (\vec \xi _{\it s} )$ and $ g_{\it s} (\vec \xi _{\it s} )$ are $|{\it s}|$-variable functions, and the decomposition (\ref{eq:ANOVA}) has $2^d$ terms in total.
\end{itemize}

\begin{example}
Consider $y=g(\vec \xi)=g(\xi_1,\xi_2)$. Since ${\cal I}=\left\{ 1, 2\right\}$, its subset includes $\emptyset$, $\left\{ 1\right\}$, $\left\{ 2\right\}$ and $\left\{ 1,2\right\}$. As a result, there exist four terms in the ANOVA decomposition (\ref{eq:ANOVA}):
\begin{itemize}
	\item for $s=\emptyset$, $g_{\emptyset}(\vec \xi_{\emptyset})=\mathbb{E}\left( g(\vec \xi)\right)=g_0$ is a constant;
	\item for $s=\left \{1 \right \}$, $g_{\left\{ 1\right\}}( \xi_1)$$=$$\hat g_{\left\{ 1\right\}}( \xi_1)-g_0$, and $\hat g_{\left\{ 1\right\}}( \xi_1)= \int\limits_{\Omega_2}  {g( {\vec \xi })\rho_2(\xi_2)d { \xi_2 } }$ is a univariate function of $\xi_1$;
	\item for $s=\left \{2 \right \}$, $g_{\left\{ 2\right\}}( \xi_2)$$=$$\hat g_{\left\{ 2\right\}}( \xi_2)-g_0$, and $\hat g_{\left\{ 2\right\}}( \xi_2)= \int\limits_{\Omega_1}  {g( {\vec \xi })\rho_1(\xi_1)d { \xi_1 } }$ is a univariate function of $\xi_2$;
	\item for $s$$=$$\left \{1,2 \right \}$,  $g_{\left\{ 1,2\right\}}( \xi_1,\xi_2)$$=$$\hat g_{\left\{ 1,2\right\}}( \xi_1,\xi_2)-g_{\left\{ 1\right\}}( \xi_1)-g_{\left\{ 2\right\}}( \xi_2)-g_0$. Since $\bar s$$ =$$\emptyset$, we have $\hat g_{\left\{ 1,2\right\}}( \xi_1,\xi_2)=g(\vec \xi)$, which is a bi-variate function.
\end{itemize}
\end{example}

Since all terms in the ANOVA decomposition are mutually orthogonal~\cite{ANOVA_sobol:2001,HDMR:1999}, we have 
\begin{align}
\mathbf{Var}\left( {g (\vec \xi  )} \right) = \sum\limits_{{\it s} \subseteq {\cal I}} {\mathbf{Var}\left( {g_{\it s} (\vec \xi _{\it s} )} \right)} 
\end{align}
where $\mathbf{Var}(\bullet)$ denotes the variance over the whole parameter space $\Omega$. What makes ANOVA practically useful is that for many engineering problems, $g(\vec \xi)$ is mainly influenced by the terms that depend only on a small number of variables, and thus it can be well approximated by a truncated ANOVA decomposition
\begin{equation}
\label{eq:ANOVA_approx}
g(\vec \xi ) \approx \sum\limits_{|{\it s}| \leq d_{\rm eff}} {g_{\it s} (\vec \xi _{\it s} )}, \;{\it s} \subseteq {\cal I}
\end{equation}
where $d_{\rm eff}\ll d$ is called the \textbf{effective dimension}.

\begin{example}
Consider $y=g(\vec \xi)$ with $d=20$. In the full ANOVA decomposition (\ref{eq:ANOVA}), we need to compute over $10^6$ terms, which is prohibitively expensive. However, if we set $d_{\rm eff}=2$, we have the following approximation
\begin{equation}
\label{eq:ANOVA_approx_example}
g(\vec \xi ) \approx g_0+\sum\limits_{j=1}^{20} {g_{{j}} (\xi_j )}+ \sum\limits_{1\leq j<k\leq 20} {g_{{j,k}} (\xi_j,\xi_k)}
\end{equation}
which contains only $221$ terms.
\end{example}

Unfortunately, it is still expensive to obtain the truncated ANOVA decomposition (\ref{eq:ANOVA_approx}) due to two reasons. First, the high-dimensional integrals in (\ref{eq:ANOVA_term}) are expensive to compute. Second, the truncated ANOVA decomposition (\ref{eq:ANOVA_approx}) still contains lots of terms when $d$ is large. In the following, we introduce anchored ANOVA that solves the first problem. The second issue will be addressed in Section~\ref{sec:adapt_anova}.

\subsubsection{Anchored ANOVA} In order to avoid the expensive multidimensional integral computation, \cite{HDMR:1999} has proposed an efficient algorithm which is called anchored ANOVA in~\cite{anchor_ANOVA_xiu:2012,anchor_ANOVA_Griebel:2010,ANOVA_zqzhang:2012}. Assuming that $\xi_k$'s have standard uniform distributions, anchored ANOVA first chooses a deterministic point called anchored point $\vec q=[q_1,\cdots, q_d] \in [0,1]^d$, and then replaces the Lebesgue measure with the Dirac measure  
\begin{equation}
\label{eq:Dirac}
d\mu ( {\vec \xi _{\bar s} } ) = \prod\limits_{k \in \bar s} {\left( {\delta \left( {\xi _k-q_k } \right)d\xi _k } \right)}.
\end{equation}
As a result, $g_0=g(\vec q)$, and 
\begin{equation}
\label{anchor_term}
\hat g_{\it s} (\vec \xi _{\it s} ) =  g\left( {\tilde \xi  } \right),\;{\rm{with}}\;\tilde \xi _k  = \left\{ \begin{array}{l}
 q_k ,\;{\rm{if}}\;k \in \bar {\it s} \\ 
 \xi _k ,\;{\rm{otherwise}}. 
 \end{array} \right.
\end{equation}
Here $\tilde \xi _k$ denotes the $k$-th element of $\tilde \xi \in \mathbb{R}^d$, $q_k$ is a fixed deterministic value, and $\xi_k$ is a random variable. Anchored ANOVA was further extended to Gaussian random parameters in~\cite{anchor_ANOVA_Griebel:2010}. In~\cite{anchor_ANOVA_xiu:2012,ANOVA_zqzhang:2012,anchor_ANOVA_xma:2010}, this algorithm was combined with stochastic collocation to efficiently solve high-dimensional stochastic partial differential equations.

\begin{example}
Consider $y$$=$$g(\xi_1,\xi_2)$. With an anchored point $\vec q=[q_1,q_2]$, we have $g_0=g(q_1,q_2)$, $\hat g_{\{1\}}(\xi_1)=g(\xi_1,q_2)$, $\hat g_{\{2\}}(\xi_2)=g(q_1,\xi_2)$ and $\hat g_{\{1,2\}}(\xi_1,\xi_2)=g(\xi_1,\xi_2)$. Computing these quantities does not involve any high-dimensional integrations.
\end{example}

\subsection{Adaptive Anchored ANOVA for Circuit/MEMS Problems}
\label{sec:adapt_anova}
\subsubsection{Extension to General Cases} In many circuit and MEMS problems, the process variations can be non-uniform and non-Gaussian. We show that anchored ANOVA can be applied to such general cases. 

{\it Observation: The anchored ANOVA in~\cite{HDMR:1999} can be applied if $\rho_k(\xi_k)>0$ for any $\xi_k\in \Omega _k$.}
\begin{proof} Let $u_k$ denote the cumulative density function for $\xi_k$, then $u_k$ can be treated as a random variable uniformly distributed on $[0,1]$. Since $\rho_k(\xi_k)>0$ for any $\xi_k \in \Omega_k$, there exists $\xi_k=\lambda_k(u_k)$ which maps $u_k$ to $\xi_k$. Therefore, $g(\xi_1,\cdots, \xi_d)=g\left(\lambda_1(u_1),\cdots, \lambda_d(u_d)\right)=\psi(\vec u)$ with $\vec u=[u_1,\cdots, u_d]$. Following (\ref{anchor_term}), we have 
\begin{equation}
\hat {\psi} _{\it s} (\vec u_{\it s} ) = \psi \left( {\tilde u } \right),\;{\rm{with}}\;\tilde u_k  = \left\{ \begin{array}{l}
 p_k ,\;{\rm{if}}\;k \in \bar {\it s} \\ 
 u_k ,\;{\rm{otherwise}},  
 \end{array} \right.
\end{equation}
where $\vec p=[p_1,\cdots, p_d]$ is the anchor point for $\vec u$. The above result can be rewritten as
\begin{equation}
\hat g_{\it s}  (\vec \xi _{\it s}  ) = g\left( {\tilde \xi   } \right){\rm{,}}\;{\rm{with}}\;\tilde \xi _k  = \left\{ \begin{array}{l}
 \lambda _k (q_k ),\;{\rm{if}}\;k \in \bar {\it s}  \\ 
 \lambda _k (\xi _k ),\;{\rm{otherwise}}, \\ 
 \end{array} \right.
\end{equation}
from which we can obtain  $g_{\it s}  (\vec \xi _{\it s}  )$ defined in (\ref{eq:ANOVA_term}). Consequently, the decomposition for $g(\vec \xi)$ can be obtained by using $\vec q=[\lambda_1(p_1),\cdots, \lambda_d(p_d)]$ as an anchor point of $\vec \xi$.
\end{proof}

\textbf{Anchor point selection.} It is is important to select a proper anchor point~\cite{ANOVA_zqzhang:2012}. In circuit and MEMS applications, we find that $\vec q=\mathbb{E}(\vec \xi)$ is a good choice.

\subsubsection{Adaptive Implementation} In order to further reduce the computational cost, the truncated ANOVA decomposition (\ref{eq:ANOVA_approx}) can be implemented in an adaptive way. Specifically, in practical computation we can ignore those terms that have small variance values. Such a treatment can produce a highly sparse generalized polynomial-chaos expansion. 

For a given effective dimension $d_{\rm eff}\ll d$, let 
\begin{equation}
{\cal S}_k=\left \{{\it s}| {\it s}\subset {\cal I}, |{\it s}|=k\right \}, \; k=1,\cdots d_{\rm eff}
\end{equation}
contain the initialized index sets for all $k$-variate terms in the ANOVA decomposition. Given an anchor point $\vec q$ and a threshold $\sigma$, starting from $k$$=$$1$, the main procedures of our ANOVA-based stochastic simulator are summarized below:
\begin{enumerate}
	\item Compute $g_0$, which is a deterministic evaluation;
	\item For every ${\it s}\in {\cal S}_k$, compute the low-dimensional function $g_{\it s}(\vec \xi_{\it s})$ by stochastic testing. The importance of $g_{\it s}(\vec \xi_{\it s})$ is measured as 
	\begin{equation}
\theta _{\it s}  = \frac{{\mathbf{Var}\left( {g_{\it s} \left( {\vec \xi _{\it s} } \right)} \right)}}{{\sum\limits_{j = 1}^k {\sum\limits_{\tilde {\it s} \in S_j } {\mathbf{Var}\left( {g_{\tilde {\it s}} \left( {\vec \xi _{\tilde {\it s}} } \right)} \right)} } }}.
	\end{equation}
	\item Update the index sets if $\theta _{\it s}<\sigma$ for ${\it s}\in {\cal S}_k$. Specifically, for $k<j\leq d_{\rm eff}$, we check its index set ${\it s}' \in {\cal S}_j$. If $s'$ contains all elements of ${\it s}$, then we remove $s'$ from ${\cal S}_j$. Once $s'$ is removed, we do not need to evaluate $g_{{\it s}'}(\vec \xi_{{\it s}'})$ in the subsequent computation.
	\item Set $k$$=k+1$, and repeat steps 2) and 3) until $k=d_{\rm def}$. 
\end{enumerate}

\begin{example}
Let $y$$=$$g(\vec \xi)$, $\vec \xi \in \mathbb{R}^{20}$ and $d_{\rm eff}=2$. Anchored ANOVA starts with 
\begin{equation}
{\cal S}_1  = \left\{ {\left\{ j \right\}} \right\}_{j=1,\cdots, 20}\;{\rm{and}}\;{\cal S}_2  = \left\{ {\left\{ {j,k} \right\}} \right\}_{1\leq j<k\leq 20}. \nonumber
\end{equation}
For $k$$=$$1$, we first utilize stochastic testing to calculate $g_{\it s}(\vec \xi_{\it s})$ and $\theta_{\it s}$ for every ${\it s} \in {\cal S}_1$. Assume 
\begin{equation}
\theta _{\left\{ 1 \right\}} >
\sigma,\;\theta _{\left\{ 2 \right\}}  >\sigma,\;{\rm{and}}\;\theta _{\left\{ j \right\}}<\sigma \; {\rm for}\; {\rm all} \; j>2, \nonumber
\end{equation}
implying that only the first two parameters are important to the output. Then, we only consider the coupling of $\xi_1$ and $\xi_2$ in ${\cal S}_2$, leading to 
\begin{equation}
{\cal S}_2  = \left\{ \left\{ {1,2} \right\} \right\}. \nonumber
\end{equation}
Consequently, for $k=2$ we only need to calculate one bi-variate function $g_{\{1,2\}}(\xi_1,\xi_2)$, yielding
\begin{equation}
\begin{array}{l}
 g\left( {\vec \xi } \right) \approx g_0  + \sum\limits_{s \in S_1 } {g_s \left( {\vec \xi _s } \right)}  + \sum\limits_{s \in S_2 } {g_s \left( {\vec \xi _s } \right)}  \\ 
 \;\;\;\;\;\;\;\; = g_0  + \sum\limits_{j = 1}^{20} {g_{\left\{ j \right\}} \left( {\xi _j } \right)}  + g_{\left\{ {1,2} \right\}} \left( {\xi _1 ,\xi _2 } \right). 
 \end{array} \nonumber
\end{equation}
\end{example}

\begin{algorithm}[t]
\caption{Stochastic Testing Circuit/MEMS Simulator Based on Adaptive Anchored ANOVA.}
\label{alg:ANOVA}
\begin{algorithmic}[1]
\STATE {Initialize ${\cal S}_k$'s and set $\beta=0$;}
\STATE {At the anchor point, run a deterministic circuit/MEMS simulation to obtain $g_0$, and set $y=g_0$;}
\STATE {\textbf{for} $k=1,\;\cdots$, $d_{\rm eff}$ \textbf{do}}
 \STATE {\hspace{10pt} {\textbf{for} each ${\it s} \in {\cal S}_k$ \textbf{do}}}
  \STATE {\hspace{20pt} run stochastic testing simulator to get the generalized \\ \hspace{20pt} polynomial-chaos expansion of $\hat {g}_{\it s}(\vec \xi_{\it s})$ };
  \STATE {\hspace{20pt} get the generalized polynomial-chaos expansion of \\ \hspace{20pt} ${g}_{\it s}(\vec \xi_{\it s})$ according to (\ref{eq:ANOVA_term})};
  \STATE {\hspace{20pt} update $\beta=\beta+\mathbf{Var}\left( {g}_{\it s}(\vec \xi_{\it s})\right)$};
  \STATE {\hspace{20pt} update $y=y+g_{\it s}(\vec \xi_{\it s})$};
 \STATE {\hspace{10pt} \textbf{end for}}
 
 \STATE {\hspace{10pt} {\textbf{for} each ${\it s} \in {\cal S}_k$ \textbf{do} }} 
  \STATE {\hspace{20pt} $\theta_{\it s}=\mathbf{Var}\left( {g}_{\it s}(\vec \xi_{\it s})\right)/ \beta;$}
  
  \STATE {\hspace{20pt} \textbf{if} $\theta_{\it s}<\sigma$}
   \STATE {\hspace{30pt} for any index set ${\it s}' \in {\cal S}_j$ with $j>k$, remove \\ \hspace{30pt} ${\it s}'$ from ${\cal S}_j$ if  ${\it s} \subset {\it s}'$}.
     \STATE {\hspace{20pt} \textbf{end if}}
 \STATE {\hspace{10pt} \textbf{end for}}
     \STATE {\textbf{end for} } 
\end{algorithmic}
\end{algorithm}

The pseudo codes of our implementation are summarized in Alg.~\ref{alg:ANOVA}. Lines $10$ to $15$ shows how to adaptively select the index sets. Let the final size of ${\cal S}_k$ be $|{\cal S}_k|$ and the total polynomial order in the stochastic testing simulator be $p$,  then the total number of samples used in Alg.~\ref{alg:ANOVA} is
\begin{equation}
N = 1 + \sum\limits_{k = 1}^{d_{\rm eff}} {|{\cal S}_k| \frac{{\left( {k + p} \right)!}}{{k!p!}}}. 
\end{equation}
Note that all univariate terms in ANOVA (i.e., $|{\it s}|=1$) are kept in our implementation. For most circuit and MEMS problems, setting the effective dimension as $2$ or $3$ can achieve a high accuracy due to the weak couplings among different random parameters. For many cases, the univariate terms dominate the output of interest, leading to a near-linear complexity with respect to the parameter dimensionality $d$. 

\textbf{Remarks.} Anchored ANOVA works very well for a large class of MEMS and circuit problems. However, in practice we also find a small number of examples (e.g., CMOS ring oscillators) that cannot be solved efficiently by the proposed algorithm, since many random variables affect significantly the output of interest. For such problems, it is possible to reduce the number of dominant random variables by a linear transform~\cite{paul:active2013} before applying anchored ANOVA. Other techniques such as compressed sensing can also be utilized to extract highly sparse surrogate models~\cite{xli2010,yxiu:2013,JPeng:2014,Hampton:2015} in the low-level simulation of our proposed hierarchical framework.

\subsubsection{Global Sensitivity Analysis}
Since each term $g_{\it s}(s_{\it s})$ is computed by stochastic testing, Algorithm~\ref{alg:ANOVA} provides a sparse generalized polynomial-chaos expansion for the output of interest: $y $$= $$\sum\limits_{|\vec \alpha | \le p} {y_{\vec \alpha } H_{\vec \alpha } (\vec \xi )} $, where most coefficients are zero. From this result, we can identify how much each parameter contributes to the output by global sensitivity analysis. Two kinds of sensitivity information can be used to measure the importance of parameter $\xi_k$: the main sensitivity $S_k$ and total sensitivity $T_k$, as computed below:
\begin{equation}
S_k  = \frac{{\sum\limits_{\alpha _k  \ne 0,\alpha _{j \ne k}  = 0\;} {\left| {y_{\vec \alpha } } \right|^2 } }}{{\mathbf{Var}(y)}},\;\;T_k  = \frac{{\sum\limits_{\alpha _k  \ne 0\;} {\left| {y_{\vec \alpha } } \right|^2 } }}{{\mathbf{Var}(y)}}.
\end{equation}

\section{Enabling High-Level Simulation by Tensor-Train Decomposition}
In this section, we show how to accelerate the high-level non-Monte-Carlo simulation by handling the obtained high-dimensional surrogate models with tensor-train decomposition~\cite{Ivan:tt_2011,Ivan:tt_svd, Ivan:tt_across}.

\subsection{Tensor-Based Three-Term Recurrence Relation}
In order to obtain the orthonormal polynomials and Gauss quadrature points/weights of $\zeta$, we must implement the three-term recurrence relation in (\ref{recurrence}). The main bottleneck is to compute the integrals in (\ref{int_cal}), since the probability density function of $\zeta$ is unknown. 

For simplicity, we rewrite the integrals in (\ref{int_cal}) as $\mathbb{E}(q(\zeta))$, with $q(\zeta)=\phi_j^2(\zeta)$ or $q(\zeta)=\zeta \phi_j^2(\zeta)$. Since the probability density function of $\zeta$ is not given, we compute the integral in the parameter space $\Omega$:
\begin{equation}
\label{eq:md_int}
\mathbb{E}\left( {q\left( \zeta  \right)} \right) = \int\limits_\Omega  {q\left( {f\left( {\vec \xi } \right)} \right)\rho (\vec \xi)d\xi _1  \cdots d\xi _d },
\end{equation}
where $f(\vec \xi)$ is a sparse generalized polynomial-chaos expansion for $\zeta$ obtained by
\begin{equation}
\zeta=f(\vec \xi)=\frac{\left(y-\mathbb{E}(y)\right)}{\sqrt{\mathbf{Var}(y)}}=\sum\limits_{|\vec \alpha | \le p} {\hat {y}_{\vec \alpha } H_{\vec \alpha } (\vec \xi )}.
\end{equation}

We compute the integral in (\ref{eq:md_int}) with the following steps:
\begin{enumerate}
	\item We utilize a multi-dimensional Gauss quadrature rule:
\begin{equation}
\label{eq:gauss_md}
\mathbb{E}\left( {q\left( \zeta  \right)} \right) \approx \sum\limits_{i_1  = 1}^{m_1} \cdots \sum\limits_{i_d  = 1}^{m_d} { {q\left( {f\left( {\xi _1^{i_1 } , \cdots ,\xi _d^{i_d } } \right)} \right)\prod\limits_{k = 1}^d {w_k^{i_k } } } } 
\end{equation}
where $m_k$ is the number of quadrature points for $\xi_k$, $(\xi_k^{i_k},w_k^{i_k})$ denotes the $i_k$-th Gauss quadrature point and weight.
\item We define two $d$-mode tensors $\mathbfcal{Q}$, $\mathbfcal{W}\in \mathbb{R}^{m_1\times m_2\cdots \times m_d}$, with each element defined as
\begin{equation}
\begin{array}{l}
 \mathbfcal{Q}\left( {i_1 , \cdots i_d } \right) = q\left( {f\left( {\xi _1^{i_1 } , \cdots ,\xi _d^{i_d } } \right)} \right),\; \\ 
 \mathbfcal{W}\left( {i_1 , \cdots i_d } \right) = \prod\limits_{k = 1}^d {w_k^{i_k } } , 
 \end{array}
\end{equation}
for $1\leq i_k\leq m_k$. Now we can rewrite (\ref{eq:gauss_md}) as the inner product of $\mathbfcal{Q}$ and $\mathbfcal{W}$:
\begin{equation}
\label{eq:tensor_inner}
\mathbb{E}\left( {q\left( \zeta  \right)} \right) \approx\left\langle {\mathbfcal{Q},\mathbfcal{W}} \right\rangle . 
\end{equation}
For simplicity, we set $m_k$$=$$m$ in this manuscript. 
\end{enumerate}

The cost of computing the tensors and the tensor inner product is $O(m^d)$, which becomes intractable when $d$ is large. Fortunately, both $\mathbfcal{Q}$ and $\mathbfcal{W}$ have low tensor ranks in our applications, and thus the high-dimensional integration (\ref{eq:md_int}) can be computed very efficiently in the following way: 
\begin{enumerate}
	\item \textbf{Low-rank representation of $\mathbfcal{W}$.} $\mathbfcal{W}$ can be written as a rank-1 tensor
\begin{equation}
\label{eq:cano_decomp}
\mathbfcal{W} = \mathbf{w}^{(1)}  \circ \mathbf{w}^{(2)}  \cdots  \circ \mathbf{w}^{(d)}, 
\end{equation}
where $\mathbf{w}^{(k)}=[w_k^1; \cdots; w_k^m]\in \mathbb{R}^{m\times 1}$ contains all Gauss quadrature weights for parameter $\xi_k$. Clearly, now we only need $O(md)$ memory to store $\mathbfcal{W}$. 

\item \textbf{Low-rank approximation for $\mathbfcal{Q}$.} $\mathbfcal{Q}$ can be well approximated by $\hat{\mathbfcal{Q}}$ with high accuracy in a tensor-train format~\cite{Ivan:tt_2011,Ivan:tt_svd, Ivan:tt_across}:
\begin{equation}
\label{eq:tt_decomp}
\hat{\mathbfcal{Q}}\left( {i_1 , \cdots i_d } \right) = \mathbfcal{G}_1 \left( {:,i_1 ,:} \right)\mathbfcal{G}_2 \left( {:,i_1 ,:} \right) \cdots \mathbfcal{G}_d \left( {:,i_d ,:} \right)
\end{equation}
with a pre-selected error bound $\epsilon$ such that
\begin{equation}
\left\| {\mathbfcal{Q} - \hat {\mathbfcal{Q}}} \right\|_F  \le \varepsilon \left\| \mathbfcal{Q} \right\|_F.
\end{equation}
For many circuit and MEMS problems, a tensor train with very small TT-ranks can be obtained even when $\epsilon=10^{-12}$ (which is very close to the machine precision). 

\item \textbf{Fast computation of (\ref{eq:tensor_inner}).} With the above low-rank tensor representations, the inner product in (\ref{eq:tensor_inner}) can be accurately estimated as
\begin{equation}
\label{eq:fast_tt_inner}
\left\langle {\hat{\mathbfcal{Q}},\mathbfcal{W}} \right\rangle  = \mathbf{T}_1   \cdots \mathbf{T}_d ,\;{\rm{with}}\;\mathbf{T}_k  = \sum\limits_{i_k=1 }^m {w_k^{i_k } \mathbfcal{G}_k \left( {:,i_k ,:} \right)} 
\end{equation}
Now the cost of computing the involved high-dimensional integration dramatically reduces to $O(dmr^2)$, which only linearly depends the parameter dimensionality $d$.

\end{enumerate}

\subsection{Efficient Tensor-Train Computation}
Now we discuss how to obtain a low-rank tensor train. An efficient implementation called \textbf{TT\_cross} is described in~\cite{Ivan:tt_across} and included in the public-domain MATALB package \textbf{TT\_Toolbox}~\cite{ttbox}. In \textbf{TT\_cross}, Skeleton decomposition is utilized to compress the TT-rank $r_k$ by iteratively searching a rank-$r_k$ maximum-volume submatrix when computing $\mathbfcal{G}_k$. A major advantage of \textbf{TT\_cross} is that we do not need to know $\mathbfcal{Q}$ {\it a-priori}. Instead, we only need to specify how to evaluate the element $\mathbfcal{Q}(i_1,\cdots, i_d)$ for a given index $(i_1,\cdots, i_d)$. As shown in~\cite{Ivan:tt_across}, with Skeleton decompositions a tensor-train decomposition needs $O(ldmr^2)$ element evaluations, where $l$ is the number of iterations in a Skeleton decomposition. For example, when $l=10$, $d=50$, $m=10$ and $r=4$ we may need up to $10^5$ element evaluations, which can take about one hour since each element of $\mathbfcal{Q}$ is a high-order polynomial function of many bottom-level random variables $\vec \xi$. 

In order to make the tensor-train decomposition of $\mathbfcal{Q}$ fast, we employ some tricks to evaluate more efficiently each element of $\mathbfcal{Q}$. The details are given below.
\begin{itemize}
	\item \textbf{Fast evaluation of} $\mathbfcal{Q}(i_1, \cdots, i_d)$. In order to reduce the cost of evaluating $\mathbfcal{Q}(i_1,\cdots, i_d)$, we first construct a low-rank tensor train $\hat{\mathbfcal{A}}$ for the intermediate-level random parameter $\zeta$, such that
\begin{equation}
\left\| {\mathbfcal{A} - \hat {\mathbfcal{A}}} \right\|_F  \le \varepsilon \left\| \mathbfcal{A} \right\|_F ,\;\mathbfcal{A}\left( {i_1 , \cdots ,i_d } \right) = f\left( {\xi _1^{i_1 } , \cdots ,\xi _d^{i_d } } \right). \nonumber
\end{equation}
Once $\hat{\mathbfcal{A}}$ is obtained, $\mathbfcal{Q}(i_1,\cdots, i_d)$ can be evaluated by
\begin{equation}
\label{tensor_element_Q}
\mathbfcal{Q}\left( {i_1 , \cdots ,i_d } \right) \approx q\left( {\hat {\mathbfcal{A}}\left( {i_1 , \cdots ,i_d } \right)} \right),
\end{equation}
which reduces to a cheap low-order univariate polynomial evaluation. However, computing $\hat{\mathbfcal{A}}(i_1,\cdots, i_d)$ by directly evaluating $\mathbfcal{A}(i_1,\cdots, i_d)$ in \textbf{TT\_cross} can be time-consuming, since $\zeta=f(\vec \xi)$ involves many multivariate basis functions.

\item \textbf{Fast evaluation of} $\mathbfcal{A}(i_1, \cdots, i_d)$. The evaluation of $\mathbfcal{A}\left( {i_1 , \cdots ,i_d } \right)$ can also be accelerated by exploiting the special structure of $f(\vec \xi)$. It is known that the generalized polynomial-chaos basis of $\vec \xi$ is 
\begin{equation}
H_{\vec \alpha } \left( {\vec \xi } \right) = \prod\limits_{k = 1}^d {\varphi _{\alpha _k }^{(k)} \left( {\xi _k } \right)}, \; \vec \alpha=[\alpha_1,\cdots, \alpha_d] 
\end{equation}
where $\varphi _{\alpha _k }^{(k)} \left( {\xi _k } \right)$ is the degree-$\alpha_k$ orthonormal polynomial of $\xi_k$, with $0\leq \alpha_k \leq p$. We first construct a $3$-mode tensor $\mathbfcal{X}\in \mathbb{R}^{d\times (p+1)\times m}$ indexed by $(k, \alpha_k +1, i_k)$ with
\begin{equation}
\label{tensor_x}
\mathbfcal{X}\left( {k,\alpha _k +1 ,i_k } \right) = \varphi _{\alpha _k }^{(k)} \left( {\xi _k^{i_k } } \right)
\end{equation}
where $\xi _k^{i_k }$ is the $i_k$-th Gauss quadrature point for parameter $\xi_k$ [as also used in (\ref{eq:gauss_md})]. Then, each element of $\mathbfcal{A}\left( {i_1 , \cdots ,i_d } \right)$ can be calculated efficiently as
\begin{equation}
\label{tensor_element_A}
\mathbfcal{A}\left( {i_1 , \cdots ,i_d } \right) = \sum\limits_{\left| {\vec \alpha } \right| < p} {\vec y_{\vec \alpha } \prod\limits_{k = 1}^d {\mathbfcal{X}\left( {k,\alpha _k+1 ,i_k } \right)} } 
\end{equation}
without evaluating the multivariate polynomials. Constructing $\mathbfcal{X}$ does not necessarily need $d(p+1)m$ polynomial evaluations, since the matrix $\mathbfcal{X}\left( {k,: ,: } \right)$ can be reused for any other parameter $\xi_j$ that has the same type of distribution with $\xi_k$.
\end{itemize}

In summary, we compute a tensor-train decomposition for $\mathbfcal{Q}$ as follows: 1) we construct the $3$-mode tensor $\mathbfcal{X}$ defined in (\ref{tensor_x}); 2) we call \textbf{TT\_cross} to compute $\hat{\mathbfcal{A}}$ as a tensor-train decomposition of $\mathbfcal{A}$, where (\ref{tensor_element_A}) is used for fast element evaluation; 3) we call \textbf{TT\_cross} again to compute $\hat{\mathbfcal{Q}}$, where (\ref{tensor_element_Q}) is used for the fast element evaluation of $\mathbfcal{Q}$. With the above fast tensor element evaluations, the computation time of $\textbf{TT\_cross}$ can be reduced from dozens of minutes to several seconds to generate some accurate low-rank tensor trains for our high-dimensional surrogate models.

\subsection{Algorithm Summary}
\begin{algorithm}[t]
\caption{Tensor-based generalized polynomial-chaos basis and Gauss quadrature rule construction for $\zeta$.}
\label{alg:tt_gauss}
\begin{algorithmic}[1]
\STATE {Initialize:  $\phi_0(\zeta)=\pi_0(\zeta)=1$, $\phi_1(\zeta)=\pi_1(\zeta)=\zeta$, $\kappa_0=\kappa_1=1$, $\gamma_0=0$, $a=1$;}
\STATE {Compute a low-rank tensor train $\hat{\mathbfcal{A}}$ for $\zeta$;}
\STATE {Compute a low-rank tensor train $\hat{\mathbfcal{Q}}$ for $q(\zeta)=\zeta^3$, and obtain $\gamma_1=\left\langle \hat{\mathbfcal{Q}},\mathbfcal{W}\right\rangle $ via (\ref{eq:fast_tt_inner});}
\STATE {\textbf{for} $j=2,\;\cdots$, $p$ \textbf{do}}
  \STATE {\hspace{10pt} get $\pi_j(\zeta)=(\zeta-\gamma_{j-1})\pi_{j-1}(\zeta)-\kappa_{j-1}\pi_{j-2}(\zeta)$ };
   \STATE {\hspace{10pt} construct a low-rank tensor train $\hat{\mathbfcal{Q}}$ for $q(\zeta)=\pi_j^2(\zeta)$,\\ \hspace{10pt} and compute $\hat a=\left\langle \hat{\mathbfcal{Q}},\mathbfcal{W}\right\rangle $ via (\ref{eq:fast_tt_inner}) };
   \STATE {\hspace{10pt} $\kappa_j=\hat a /a$, and update $a=\hat a$ };
   \STATE {\hspace{10pt} construct a low-rank tensor train $\hat{\mathbfcal{Q}}$ for $q(\zeta)=\zeta\pi_j^2(\zeta)$,\\ \hspace{10pt} and compute $\gamma_j= \left\langle \hat{\mathbfcal{Q}},\mathbfcal{W}\right\rangle/a$ };
     \STATE {\hspace{10pt} normalization: $\phi_j(\zeta)=\frac{\pi_j(\zeta)}{\sqrt{\kappa_0\cdots \kappa_j}}$ };
\STATE {\textbf{end for} } 
\STATE {Form matrix $\mathbf{J}$ in (\ref{eq:jmatrix}); }
\STATE {Eigenvalue decomposition: $\mathbf{J} = \mathbf{U}\Sigma \mathbf{U}^T$ ;}
\STATE {Compute the Gauss-quadrature abscissa $\zeta^j=\Sigma(j,j)$ and weight $w^j=\left(\mathbf{U}(1,j)\right)^2$ for $j=1,\cdots, p+1$ ;}
\end{algorithmic}
\end{algorithm}
Given the Gauss quadrature rule for each bottom-level random parameter $\xi_k$, our tensor-based three-term recurrence relation for an intermediate-level random parameter $\zeta$ is summarized in Alg.~\ref{alg:tt_gauss}. This procedure can be repeated for all $\zeta_i$'s to obtain their univariate generalized polynomial-chaos basis functions and Gauss quadrature rules, and then the stochastic testing simulator~\cite{zzhang:tcad2013,zzhang:tcas2_2013,zzhang:iccad_2013} (and any other standard stochastic spectral method~\cite{col:2005,sfem,gPC2002}) can be employed to perform high-level stochastic simulation.

\textbf{Remarks.} 1) If the outputs of a group of subsystems are identically independent, we only need to run Alg.~\ref{alg:tt_gauss} once and reuse the results for the other subsystems in the group. 2) When there exist many subsystems, our ANOVA-based stochastic solver may also be utilized to accelerate the high-level simulation.

\section{Numerical Results}

In this section, we verify the proposed algorithm on a MEMS/IC co-design example with high-dimensional random parameters. All simulations are run in MATLAB and executed on a 2.4GHz laptop with 4GB memory.
\subsection{MEMS/IC Example}
\begin{figure}[t]
	\centering
		\includegraphics[width=2.8in]{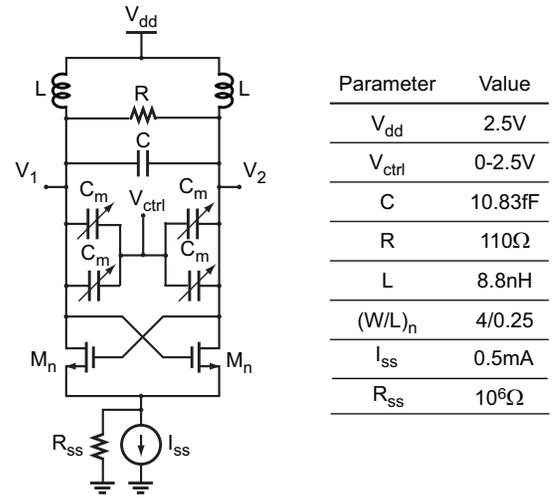} 
\caption{Schematic of the oscillator circuit with $4$ MEMS capacitors (denoted as $C_{\rm m}$), with $184$ random parameters in total.}
	\label{fig:mems_vco_high}
\end{figure}

In order to demonstrate the application of our hierarchical uncertainty quantification in high-dimensional problems, we consider the oscillator circuit shown in Fig.~\ref{fig:mems_vco_high}. This oscillator has four identical RF MEMS switches acting as tunable capacitors. The MEMS device used in this paper is a prototyping model of the RF MEMS capacitor reported in~\cite{Dana:2011,Stamper:2011}. 
\begin{figure}[t]
	\centering
		\includegraphics[width=3.3in]{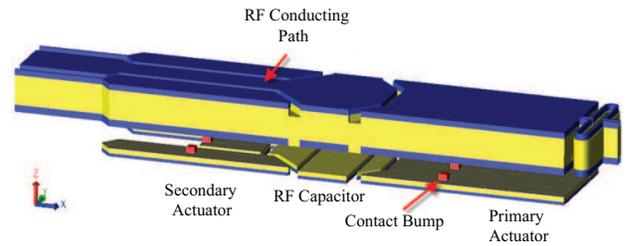} 
\caption{3-D schematic of the RF MEMS capacitor.}
	\label{fig:RF_cap}
\end{figure}

Since the MEMS switch has a symmetric structure, we construct a model for only half of the design, as shown in Fig.~\ref{fig:RF_cap}. The simulation and measurement results in~\cite{zzhang:JMEMS2014} show that the pull-in voltage of this MEMS switch is about $37$ V. When the control voltage is far below the pull-in voltage, the MEMS capacitance is small and almost constant. In this paper, we set the control voltage to $2.5$ V, and thus the MEMS switch can be regarded as a small linear capacitor. As already shown in~\cite{Matt:2013}, the performance of this MEMS switch can be influenced significantly by process variations. 

In our numerical experiments, we use $46$ independent random parameters with Gaussian and Gamma distributions to describe the material (e.g, conductivity and dielectric constants), geometric (e.g., thickness of each layer, width  and length of each mechanical component) and environmental (e.g., temperature) uncertainties of each switch. For each random parameter, we assume that its standard deviation is $3\%$ of its mean value. In the whole circuit, we have $184$ random parameters in total. Due to such high dimensionality, simulating this circuit by stochastic spectral methods is a challenging task.
\begin{table}[t]
	\centering 
	\caption{Different hierarchical simulation methods.}	
	\label{tab:method_list}
	\begin{threeparttable}
	\begin{tabular}{|c||c|c|}
	\hline%\hline		
Method& Low-level simulation& High-level simulation\\
\thickhline	
	Proposed  & Alg. 1 & stochastic testing~\cite{zzhang:tcas2_2013} \\ \hline
	Method 1~\cite{Felt:1996} & Monte Carlo & Monte Carlo \\ \hline
	Method 2  & Alg. 1 & Monte Carlo   \\	
\hline%\hline
	\end{tabular} 	
 \end{threeparttable}	 	
\end{table}

\begin{figure}[t]
	\centering
		\includegraphics[width=3.3in]{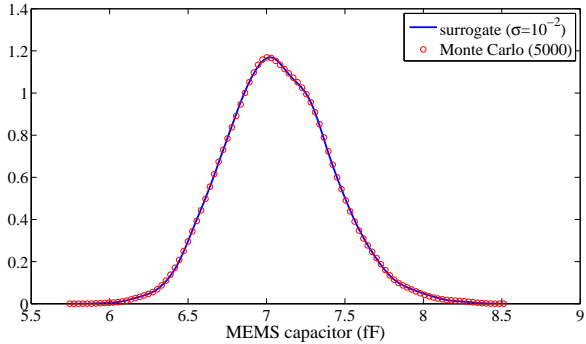} 
\caption{Comparison of the density functions obtained by our surrogate model and by $5000$-sample Monte Carlo analysis of the original MEMS equation.}
	\label{fig:RFcap_pdf}
\end{figure}

In the following experiments, we simulate this challenging design case using our proposed hierarchical stochastic spectral methods. We also compare our algorithm with other two kinds of hierarchical approaches listed in Table~\ref{tab:method_list}. In Method 1, both low-level and high-level simulations use Monte Carlo, as suggested by~\cite{Felt:1996}. In Method 2, the low-level simulation uses our ANOVA-based sparse simulator (Alg. 1), and the high-level simulation uses Monte Carlo.
\begin{table*}[t]
	\centering 
	\caption{Surrogate model extraction with different $\sigma$ values.}	
	\label{tab:anova_converge}
	\begin{threeparttable}
	\begin{tabular}{|c||c|c|c|c|c|c|}
	\hline%\hline		
$\sigma$& $\#$ $|{\it s}|$$=1$& $\#$ $|{\it s}|$$=2$ & $\#$ $|{\it s}|$$=3$ & $\#$ ANOVA terms &$\#$ nonzero gPC terms & $\#$ samples\\
\thickhline	
	$0.5$  & $46$  & $0$ & $0$ &  $47$ & $81$ & $185$ \\ \hline
	$0.1$ to $10^{-3}$ & $46$ & $3$ & $0$ &  $50$ & $90$ & $215$ \\ \hline
	$10^{-4}$  & $46$ & $10$ & $1$ &  $58$ & $112$ & $305$ \\ \hline
	$10^{-5}$  & $46$ & $21$ & $1$ &  $69$ & $144$ & $415$  \\	
\hline%\hline
	\end{tabular} 	
 \end{threeparttable}	 	
\end{table*}

\subsection{Surrogate Model Extraction}
In order to extract an accurate surrogate model for the MEMS capacitor, Alg. 1 is implemented in the commercial network-based MEMS simulation tool MEMS+~\cite{memsp_mannual} of Coventor Inc. Each MEMS switch is described by a stochastic differential equation [c.f. (\ref{eq:sdae})] with consideration of process variations. In order to compute the MEMS capacitor, we can ignore the derivative terms and solve for the static solutions.

By setting $\sigma=10^{-2}$, our ANOVA-based stochastic MEMS simulator generates a sparse 3rd-order generalized polynomial chaos expansion with only $90$ non-zero coefficients, requiring only $215$ simulation samples and $8.5$ minutes of CPU time in total. This result has only $3$ bivariate terms and no three-variable terms in ANOVA decomposition, due to the very weak couplings among different random parameters. Setting $\sigma=10^{-2}$ can provide a highly accurate generalized polynomial chaos expansion for the MEMS capacitor, which has a relative error around $10^{-6}$ (in the ${\it L}_2$ sense) compared to that obtained by setting $\sigma=10^{-5}$. 

By evaluating the surrogate model and the original model (by simulating the original MEMS equation) with $5000$ samples, we have obtained the same probability density curves shown in Fig.~\ref{fig:RFcap_pdf}. Note that using the standard stochastic testing simulator~\cite{zzhang:tcad2013,zzhang:tcas2_2013,zzhang:iccad_2013} requires $18424$ basis functions and simulation samples for this high-dimensional example, which is prohibitively expensive on a regular computer. When the effective dimension $d_{\rm eff}$ is set as $3$, there should be $16262$ terms in the truncated ANOVA decomposition (\ref{eq:ANOVA_approx}). However, due to the weak couplings among different random parameters, only $90$ of them are non-zero.

\begin{figure}[t]
	\centering
		\includegraphics[width=3.3in]{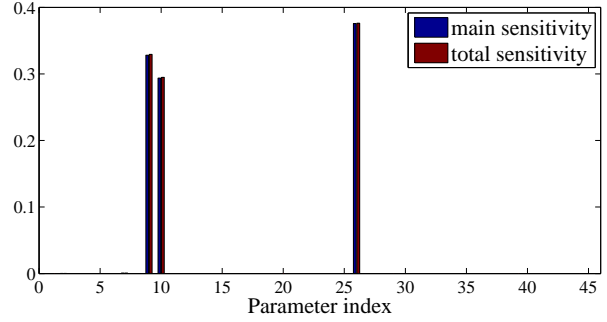} 
\caption{Main and total sensitivities of different random parameters for the RF MEMS capacitor.}
	\label{fig:RFcap_sensitivity}
\end{figure}

We can get surrogate models with different accuracies by changing the threshold $\sigma$. Table~\ref{tab:anova_converge} has listed the number of obtained ANOVA terms, the number of non-zero generalized polynomial chaos (gPC) terms and the number of required simulation samples for different values of $\sigma$. From this table, we have the following observations:
\begin{enumerate}
	\item When $\sigma$ is large, only $46$ univariate terms (i.e., the terms with $|{\it s}|=1$) are obtained. This is because the variance of all univariate terms are regarded as small, and thus all multivariate terms are ignored.
	\item When $\sigma$ is reduced (for example, to $0.1$), three dominant bivariate terms (with $|{\it s}|=2$) are included by considering the coupling effects of the three most influential random parameters. Since the contributions of other parameters are insignificant, the result does not change even if $\sigma$ is further decreased to $10^{-3}$.
	\item A three-variable term (with $|{\it s}|=3$) and some bivariate coupling terms among other parameters can only be captured when $\sigma$ is reduced to $10^{-4}$ or below. In this case, the effect of some non-dominant parameters can be captured.
\end{enumerate}
\begin{figure}[t]
	\centering
		\includegraphics[width=3.3in]{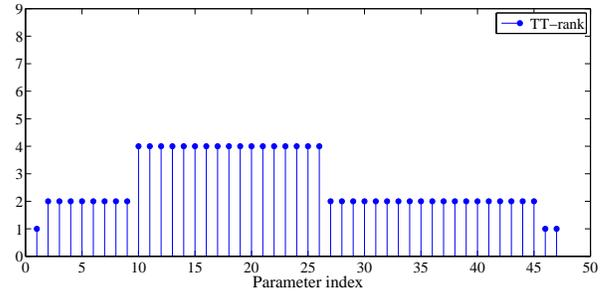} 
\caption{TT-rank for the surrogate model of the RF MEMS capacitor.}
	\label{fig:RFCap_ttrank}
\end{figure}

\begin{figure*}[t]
	\centering
		\includegraphics[width=5.5in]{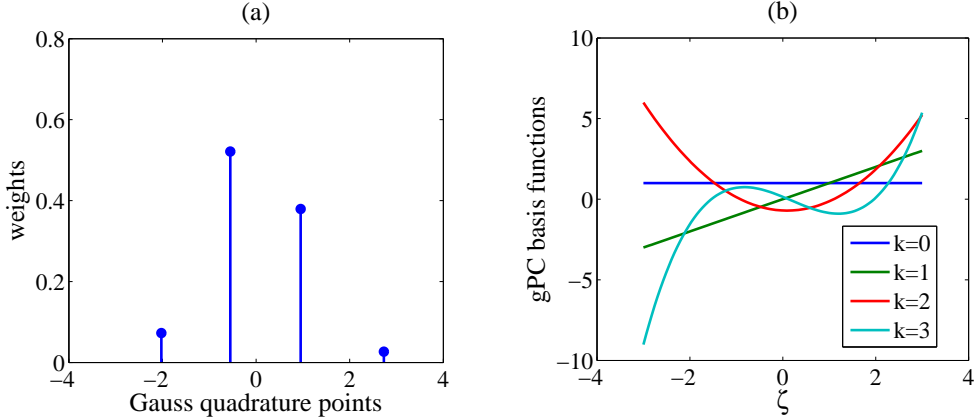} 
\caption{(a) Gauss quadrature rule and (b) generalized polynomial chaos (gPC) basis functions for the RF MEMS capacitor.}
	\label{fig:RFCap_gauss}
\end{figure*}

Fig.~\ref{fig:RFcap_sensitivity} shows the global sensitivity of this MEMS capacitor with respect to all $46$ random parameters. The output is dominated by only $3$ parameters. The other $43$ parameters contribute to only $2\%$ of the capacitor's variance, and thus their main and total sensitivities are almost invisible in Fig.~\ref{fig:RFcap_sensitivity}. This explains why the generalized polynomial-chaos expansion is highly sparse. Similar results have already been observed in the statistical analysis of CMOS analog circuits~\cite{zzhang_cicc2014}.

\subsection{High-Level Simulation}
The surrogate model obtained with $\sigma=10^{-2}$ is imported into the stochastic testing circuit simulator described in~\cite{zzhang:tcad2013,zzhang:tcas2_2013,zzhang:iccad_2013} for high-level simulation. At the high-level, we have the following differential equation to describe the oscillator:
\begin{equation}
\label{eq:sdae_high}
\begin{array}{l}
 \displaystyle{\frac{{d\vec q\left( {\vec x( {t,\vec \zeta } ),\vec \xi } \right)}}{{dt}} }+ \vec f\left( {\vec x( {t,\vec \zeta } ),\vec \zeta }, u \right) = 0 
 \end{array}
\end{equation}
where the input signal $u$ is constant, ${\vec \zeta}$=$[\zeta_1,\cdots,\zeta_4]\in\mathbb{R}^4$ are the intermediate-level random parameters describing the four MEMS capacitors. Since the oscillation period $T(\vec \zeta)$ now depends on the MEMS capacitors, the periodic steady-state can be written as $\vec x (t,\zeta)=\vec x (t+T(\vec \zeta),\zeta)$. We simulate the stochastic oscillator by the following steps~\cite{zzhang:tcas2_2013}:
\begin{enumerate}
	\item Choose a constant $T_0>0$ to define an unknown scaling factor $a(\vec \zeta)=T(\vec \zeta)/T_0$ and a scaled time axis $\tau=t/\alpha(\vec \zeta)$. With this scaling factor, we obtain a reshaped waveform $\vec z(\tau,\vec \zeta)=\vec x(t/a(\vec \zeta),\vec \zeta)$. At the steady state, we have $\vec z(\tau,\vec \zeta)=\vec z(\tau +T_0,\vec \zeta)$. In other words, the reshaped waveform has a period $T_0$ independent of $\vec \zeta$.	
	\item Rewrite (\ref{eq:sdae_high}) on the scaled time axis:
	\begin{equation}
\label{eq:sdae_high_scaled}
\begin{array}{l}
 \displaystyle{\frac{{d\vec q\left( {\vec z( {\tau,\vec \zeta } ),\vec \xi } \right)}}{{d \tau}} }+ a(\vec \zeta)\vec f\left( {\vec z( {\tau,\vec \zeta } ),\vec \zeta }, u \right) = 0 .
 \end{array}
\end{equation}
\item Approximate $\vec z( {\tau,\vec \zeta } )$ and $a(\vec \zeta)$ by generalized polynomial chaos expansions of $\vec \zeta$. Then, convert (\ref{eq:sdae_high_scaled}) to a larger-scale deterministic equation by stochastic testing. Solve the resulting deterministic equation by shooting Newton with a phase constraints, which would provide the coefficients in the generalized polynomial-chaos expansions of $\vec z( {\tau,\vec \zeta } )$ and $\alpha(\vec \zeta)$~\cite{zzhang:tcas2_2013}. 
\item Map $\vec z( {\tau,\vec \zeta } )$ to the original time axis, we obtain the periodic steady state of $\vec x (t,\zeta)$.
\end{enumerate}
\begin{figure*}[t]
	\centering
		\includegraphics[width=5.5in]{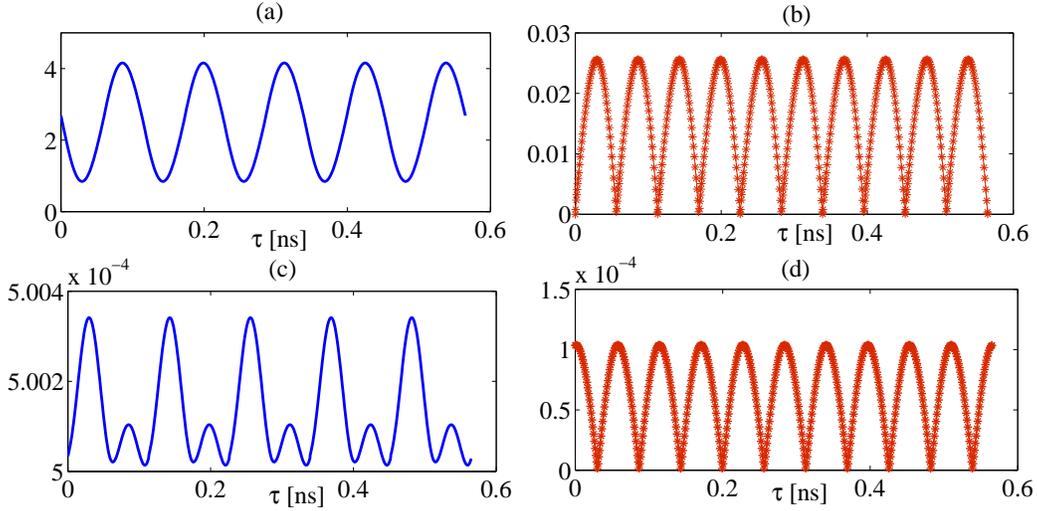} 
\caption{Simulated waveforms on the scaled time axis $\tau=t/a(\vec \zeta)$. (a) and (b): the mean and standard deviation of $V_{\rm {out}1}$ (unit: V), respectively; (c) and (d): the mean and standard deviation of the current (unit: A) from $V_{\rm dd}$, respectively.}
	\label{fig:MEMSVCO_MeanStd}
\end{figure*}

In order to apply stochastic testing in Step 3), we need to compute some specialized orthonormal polynomials and Gauss quadrature points for each intermediate-level parameter $\zeta_i$. We use $9$ quadrature points for each bottom-level parameter $\xi_k$ to evaluate the high-dimensional integrals involved in the three-term recurrence relation. This leads to $9^{46}$ function evaluations at all quadrature points, which is prohibitively expensive. 

To handle the high-dimensional MEMS surrogate models, the following tensor-based procedures are employed:
\begin{itemize}
	\item With Alg.~\ref{alg:tt_gauss}, a low-rank tensor train of $\zeta_1$ is first constructed for an MEMS capacitor. For most dimensions the rank is only $2$, and the highest rank is $4$, as shown in Fig.~\ref{fig:RFCap_ttrank}.
	\item Using the obtained tensor train, the Gauss quadrature points and generalized polynomial chaos basis functions are efficiently computed, as plotted in Fig.~\ref{fig:RFCap_gauss}.
\end{itemize}

The total CPU time for constructing the tensor trains and computing the basis functions and Gauss quadrature points/weights is about $40$ seconds in MATALB. If we directly evaluate the high-dimensional multivariate generalized polynomial-chaos expansion, the three-term recurrence relation requires almost $1$ hour. The obtained results can be reused for all MEMS capacitors since they are independently identical.

With the obtained basis functions and Gauss quadrature points/weights for each MEMS capacitor, the stochastic periodic steady-state solver~\cite{zzhang:tcas2_2013} is called at the high level to simulate the oscillator. Since there are $4$ intermediate-level parameters $\zeta_i$'s, only $35$ basis functions and testing samples are required for a $3$rd-order generalized polynomial-chaos expansion, leading to a simulation cost of only $56$ seconds in MATLAB. 

\begin{figure}[t]
	\centering
		\includegraphics[width=3.3in]{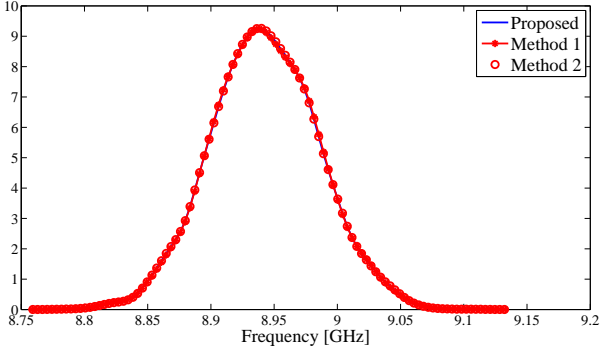} 
\caption{Probability density functions of the oscillation frequency.}
	\label{fig:MEMSVCO_freq_pdf}
\end{figure}

Fig.~\ref{fig:MEMSVCO_MeanStd} shows the waveforms from our algorithm at the scaled time axis $\tau =t/a(\vec \zeta)$. The high-level simulation generates a generalized polynomial-chaos expansion for all nodal voltages, branch currents and the exact parameter-dependent period. Evaluating the resulting generalized polynomial-chaos expansion with $5000$ samples, we have obtained the density function of the frequency, which is consistent with those from Method 1 (using $5000$ Monte Carlo samples at both levels) and Method 2 (using Alg. 1 at the low level and using $5000$ Monte-Carlo samples at the high level), as shown in Fig.~\ref{fig:MEMSVCO_freq_pdf}. 

\begin{figure}[t]
	\centering
		\includegraphics[width=3.3in]{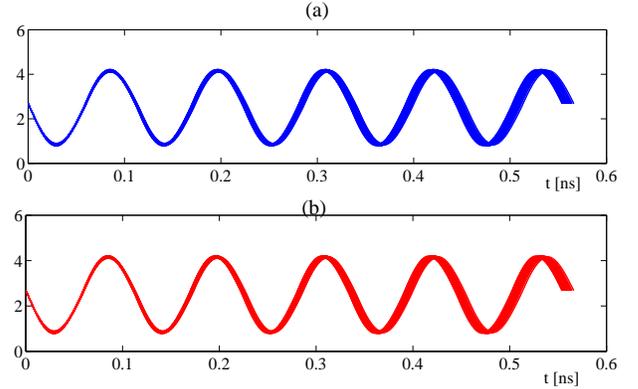} 
\caption{Realization of the output voltages (unit: volt) at $100$ bottom-level samples, generated by (a) proposed method and (b) Method 1.}
	\label{fig:wave_mc}
\end{figure}

In order to show the variations of the waveform, we further plot the output voltages for $100$ bottom-level random samples. As shown in Fig.~\ref{fig:wave_mc}, the results from our proposed method and from Method 1 are indistinguishable from each other.

\subsection{Complexity Analysis}

Table~\ref{tab:hierUQ_compare} has summarized the performances of all three methods. In all Monte Carlo analysis, $5000$ random samples are utilized. If Method 1~\cite{Felt:1996} is used, Monte Carlo has to be repeatedly used for each MEMS capacitor, leading to extremely long CPU time due to the slow convergence. If Method 2 is used, the efficiency of the low-level surrogate model extraction can be improved due to the employment of generalized polynomial-chaos expansion, but the high-level simulation is still time-consuming. Since our proposed technique utilizes fast stochastic testing algorithms at both levels, this high-dimensional example can be simulated at very low computational cost, leading to $92\times$ speedup over Method 1 and $14\times$ speedup over Method 2.
%If there exist many subsystems, our proposed ANOVA-based sparse stochastic testing simulator can also be employed to accelerate the high-level simulation.
%\begin{figure}[t]
%	\centering
%		\includegraphics[width=3.3in]{RFcap_pdf.eps} 
%\caption{Comparison of the density function of our surrogate model with the exact result.}
%	\label{fig:RFcap_pdf}
%\end{figure}
\begin{table*}[t]
	\centering 
	\caption{CPU times of different hierarchical stochastic simulation algorithms.}	
	\label{tab:hierUQ_compare}
	\begin{threeparttable}
	\scriptsize	
		\begin{tabular}{|c|c|c|c|c|c|}
	\hline%\hline	
	\multirow{2}{*}{Simulation Method}	&\multicolumn{2}{|c|}{Low level} &\multicolumn{2}{|c|}{High level}	 & \multirow{2}{*}{Total simulation cost} \\ \cline{2-5}
		& Method & CPU time & Method & CPU time  &\\ \thickhline		
	Proposed & Alg. 1 &  $8.5$ min & stochastic testing & $1.5$ minute   & Low ($10$ min)\\  \hline		
	Method 1 & Monte Carlo & $13.2$ h    & Monte Carlo & $2.2$ h  & High ($15.4$ h)\\  \hline	
		Method 2 & Alg. 1 & $8.5$ min & Monte Carlo &  $2.2$ h  & Medium ($2.3$ h)\\ \hline
	\end{tabular} 	
\end{threeparttable}	
\end{table*}

\section{Conclusions and Future Work}
This paper has proposed a framework to accelerate the hierarchical uncertainty quantification of stochastic circuits/systems with high-dimensional subsystems. We have developed an ANOVA-based stochastic testing simulator to accelerate the low-level simulation, and a tensor-based technique for handling high-dimensional surrogate models at the high level. Both algorithms have a linear (or near-linear) complexity with respect to the parameter dimensionality. Our simulator has been tested on an oscillator circuit with four MEMS capacitors and totally $184$ random parameters, achieving highly accurate results at the cost of $10$-min CPU time in MATLAB. In such example, our method is over $92\times$ faster than the hierarchical Monte Carlo method developed in~\cite{Felt:1996}, and is about $14\times$ faster than the method that uses ANOVA-based solver at the low level and Monte Carlo at the high level.

There are lots of problems worth investigation in the direction of hierarchical uncertainty quantification. Some unsolved important questions include:
\begin{enumerate}
	\item How to extract a high-dimensional surrogate model such that the tensor rank is as small as possible (or the tensor rank is below a provided upper bound)?
	\item How to perform non-Monte-Carlo hierarchical uncertainty quantification when the outputs of different blocks are correlated?
	\item How to perform non-Monte-Carlo hierarchical uncertainty quantification when $y_i$ depends on some varying variables (e.g., time and frequency)?
\end{enumerate}

\section*{Acknowledgments}
The authors would like to thank Coventor Inc. for providing the MEMS+ license and the MEMS switch design files. We would like to thank Shawn Cunningham and Dana Dereus of Wispry for providing access to the MEMS switch data. We are grateful to Dr. Giovanni Marucci for providing the oscillator design parameters, as well as Prof. Paolo Maffezzoni and Prof. Ibrahim Elfadel for their technical suggestions.
\bibliographystyle{IEEEtran}
\bibliography{date}

\begin{thebibliography}{10}
\providecommand{\url}[1]{#1}
\csname url@samestyle\endcsname
\providecommand{\newblock}{\relax}
\providecommand{\bibinfo}[2]{#2}
\providecommand{\BIBentrySTDinterwordspacing}{\spaceskip=0pt\relax}
\providecommand{\BIBentryALTinterwordstretchfactor}{4}
\providecommand{\BIBentryALTinterwordspacing}{\spaceskip=\fontdimen2\font plus
\BIBentryALTinterwordstretchfactor\fontdimen3\font minus
  \fontdimen4\font\relax}
\providecommand{\BIBforeignlanguage}[2]{{%
\expandafter\ifx\csname l@#1\endcsname\relax
\typeout{** WARNING: IEEEtran.bst: No hyphenation pattern has been}%
\typeout{** loaded for the language `#1'. Using the pattern for}%
\typeout{** the default language instead.}%
\else
\language=\csname l@#1\endcsname
\fi
#2}}
\providecommand{\BIBdecl}{\relax}
\BIBdecl

\bibitem{zzhang_cicc2014}
Z.~Zhang, X.~Yang, G.~Marucci, P.~Maffezzoni, I.~M. Elfadel, G.~Karniadakis,
  and L.~Daniel, ``Stochastic testing simulator for integrated circuits and
  {MEMS}: Hierarchical and sparse techniques,'' in \emph{Proc. IEEE Custom
  Integrated Circuits Conf.}\hskip 1em plus 0.5em minus 0.4em\relax San Jose,
  CA, Sept. 2014.

\bibitem{variation2008}
D.~S. Boning, ``Variation,'' \emph{IEEE Trans. Semiconductor Manufacturing},
  vol.~21, no.~1, pp. 63--71, Feb 2008.

\bibitem{LiYu:2014dac}
L.~Yu, S.~Saxena, C.~Hess, A.~Elfadel, D.~A. Antoniadis, and D.~S. Boning,
  ``Remembrance of transistors past: Compact model parameter extraction using
  {Bayesian} inference and incomplete new measurements,'' in \emph{Proc. Design
  Automation Conf.}\hskip 1em plus 0.5em minus 0.4em\relax San Francisco, CA,
  Jun 2014, pp. 1--6.

\bibitem{LiYu:2014date}
L.~Yu, S.~Saxena, C.~Hess, I.~M. Elfadel, D.~A. Antoniadis, and D.~S. Boning,
  ``Efficient performance estimation with very small sample size via physical
  subspace projection and maximum a posteriori estimation,'' in \emph{Proc.
  Design Automation and Test in Europe}.\hskip 1em plus 0.5em minus 0.4em\relax
  Dresden, Germany, March 2014, pp. 1--6.

\bibitem{LiYu:2013date}
L.~Yu, L.~Wei, D.~A. Antoniadis, I.~M. Elfadel, and D.~S. Boning, ``Statistical
  modeling with the virtual source {MOSFET} model,'' in \emph{Proc. Design
  Automation and Test in Europe}.\hskip 1em plus 0.5em minus 0.4em\relax
  Grenoble, France, March 2013, pp. 1454--1457.

\bibitem{LiYu:2012ISQED}
L.~Yu, W.-Y. Chang, K.~Zuo, J.~Wang, D.~Yu, and D.~S. Boning, ``Methodology for
  analysis of {TSV} stress induced transistor variation and circuit
  performance,'' in \emph{Proc. Int. Symp. Quality Electronic Design}.\hskip
  1em plus 0.5em minus 0.4em\relax Santa Clara, CA, March 2012, pp. 216--222.

\bibitem{sfem}
R.~Ghanem and P.~Spanos, \emph{Stochastic finite elements: a spectral
  approach}.\hskip 1em plus 0.5em minus 0.4em\relax Springer-Verlag, 1991.

\bibitem{gPC2002}
D.~Xiu and G.~E. Karniadakis, ``The {Wiener-Askey} polynomial chaos for
  stochastic differential equations,'' \emph{SIAM J. Sci. Comp.}, vol.~24,
  no.~2, pp. 619--644, Feb 2002.

\bibitem{col:2005}
D.~Xiu and J.~S. Hesthaven, ``High-order collocation methods for differential
  equations with random inputs,'' \emph{SIAM J. Sci. Comp.}, vol.~27, no.~3,
  pp. 1118--1139, Mar 2005.

\bibitem{Ivo:2007}
I.~Babu\v{s}ka, F.~Nobile, and R.~Tempone, ``A stochastic collocation method
  for elliptic partial differential equations with random input data,''
  \emph{SIAM J. Numer. Anal.}, vol.~45, no.~3, pp. 1005--1034, Mar 2007.

\bibitem{Nobile:2008}
F.~Nobile, R.~Tempone, and C.~G. Webster, ``A sparse grid stochastic
  collocation method for partial differential equations with random input
  data,'' \emph{SIAM J. Numer. Anal.}, vol.~46, no.~5, pp. 2309--2345, May
  2008.

\bibitem{Nobile:2008_2}
------, ``An anisotropic sparse grid stochastic collocation method for partial
  differential equations with random input data,'' \emph{SIAM J. Numer. Anal.},
  vol.~46, no.~5, pp. 2411--2442, May 2008.

\bibitem{SingheeR10}
A.~Singhee and R.~A. Rutenbar, ``Why {Q}uasi-{M}onte {C}arlo is better than
  {M}onte {C}arlo or latin hypercube sampling for statistical circuit
  analysis,'' \emph{IEEE Trans. CAD of Integr. Circuits and Syst.}, vol.~29,
  no.~11, pp. 1763--1776, Nov. 2010.

\bibitem{zzhang:tcad2013}
Z.~Zhang, T.~A. El-Moselhy, I.~M. Elfadel, and L.~Daniel, ``Stochastic testing
  method for transistor-level uncertainty quantification based on generalized
  polynomial chaos,'' \emph{IEEE Trans. CAD of Integr. Circuits and Syst.},
  vol.~32, no.~10, pp. 1533--1545, Oct 2013.

\bibitem{zzhang:tcas2_2013}
Z.~Zhang, T.~A. El-Moselhy, P.~Maffezzoni, I.~M. Elfadel, and L.~Daniel,
  ``Efficient uncertainty quantification for the periodic steady state of
  forced and autonomous circuits,'' \emph{IEEE Trans. Circuits and Systems II:
  Express Briefs}, vol.~60, no.~10, Oct 2013.

\bibitem{zzhang:iccad_2013}
Z.~Zhang, I.~M. Elfadel, and L.~Daniel, ``Uncertainty quantification for
  integrated circuits: Stochastic spectral methods,'' in \emph{Proc. Int. Conf.
  Computer-Aided Design}.\hskip 1em plus 0.5em minus 0.4em\relax San Jose, CA,
  Nov 2013, pp. 803--810.

\bibitem{Strunz:2008}
K.~Strunz and Q.~Su, ``Stochastic formulation of {SPICE}-type electronic
  circuit simulation with polynomial chaos,'' \emph{ACM Trans. Modeling and
  Computer Simulation}, vol.~18, no.~4, pp. 15:1--15:23, Sep 2008.

\bibitem{Tao:2007}
J.~Tao, X.~Zeng, W.~Cai, Y.~Su, D.~Zhou, and C.~Chiang, ``Stochastic
  sparse-grid collocation algorithm ({SSCA}) for periodic steady-state analysis
  of nonlinear system with process variations,'' in \emph{Porc. Asia and South
  Pacific Design Automation Conference}, 2007, pp. 474--479.

\bibitem{Manfredi:tcas1_2014}
P.~Manfredi, D.~V. Ginste, D.~{De Zutter}, and F.~Canavero, ``Stochastic
  modeling of nonlinear circuits via {SPICE}-compatible spectral equivalents,''
  \emph{IEEE Trans. Circuits Syst. I: Regular Papers}, 2014.

\bibitem{Pulch:2011_1}
R.~Pulch, ``Modelling and simulation of autonomous oscillators with random
  parameters,'' \emph{Mathematics and Computers in Simulation}, vol.~81, no.~6,
  pp. 1128--1143, Feb 2011.

\bibitem{Stievano:2011}
I.~S. Stievano, P.~Manfredi, and F.~G. Canavero, ``Stochastic analysis of
  multiconductor cables and interconnects,'' \emph{IEEE Trans. Electromagnetic
  Compatibility}, vol.~53, no.~2, pp. 501--507, May 2011.

\bibitem{cmpt2012}
D.~V. Ginste, D.~D. Zutter, D.~Deschrijver, T.~Dhaene, P.~Manfredi, and
  F.~Canavero, ``Stochastic modeling-based variability analysis of on-chip
  interconnects,'' \emph{IEEE Trans. Components, Packaging and Manufacturing
  Technology}, vol.~2, no.~7, pp. 1182--1192, Jul. 2012.

\bibitem{Vrudhula:2006}
S.~Vrudhula, J.~M. Wang, and P.~Ghanta, ``Hermite polynomial based interconnect
  analysis in the presence of process variations,'' \emph{IEEE Trans. CAD
  Integr. Circuits Syst.}, vol.~25, no.~10, pp. 2001--2011, Oct. 2006.

\bibitem{Tarek_ISQED:11}
T.~Moselhy and L.~Daniel, ``Variation-aware stochastic extraction with large
  parameter dimensionality: Review and comparison of state of the art intrusive
  and non-intrusive techniques,'' in \emph{Proc. Intl. Symp. Quality Electronic
  Design}, Mar. 2011, pp. 14--16.

\bibitem{Tarek_DAC:08}
------, ``Stochastic integral equation solver for efficient variation aware
  interconnect extraction,'' in \emph{Proc. Design Auto. Conf.}, Jun. 2008, pp.
  415--420.

\bibitem{sMOR2012}
P.~Sumant, H.~Wu, A.~Cangellaris, and N.~R. Aluru, ``Reduced-order models of
  finite element approximations of electromagnetic devices exhibiting
  statistical variability,'' \emph{IEEE Trans. Antennas and Propagation},
  vol.~60, no.~1, pp. 301--309, Jan. 2012.

\bibitem{MEMS_uq_jmems09}
N.~Agarwal and N.~R. Aluru, ``Stochastic analysis of electrostatic {MEMS}
  subjected to parameter variations,'' \emph{J. Microelectromech. Syst.},
  vol.~18, no.~6, pp. 1454--1468, Dec. 2009.

\bibitem{zzhang:tcad2014}
Z.~Zhang, T.~A. El-Moselhy, I.~M. Elfadel, and L.~Daniel, ``Calculation of
  generalized polynomial-chaos basis functions and {Gauss} quadrature rules in
  hierarchical uncertainty quantification,'' \emph{IEEE Trans. CAD Integr.
  Circuits Syst.}, vol.~33, no.~5, pp. 728--740, May 2014.

\bibitem{Ng:2014}
L.~W.~T. Ng and K.~E. Wilcox, ``A multi-information source approach to aircraft
  conceptual design under uncertainty,'' under review.

\bibitem{Ng:2014_opt}
------, ``Multifidelity approaches for optimization under uncertainty,''
  \emph{Int. J. Numerical Meth. Eng.}, Sept. 2014.

\bibitem{Allaire:2014}
D.~Allaire and K.~E. Wilcox, ``A mathematical and computational framework for
  multifidelity design and analysis with computer models,'' \emph{Int. J.
  Uncertainty Quantification}, vol.~4, no.~1, pp. 1--20, 2014.

\bibitem{Felt:1996}
E.~Felt, S.~Zanella, C.~Guardiani, and A.~Sangiovanni-Vincentelli,
  ``Hierarchical statistical characterization of mixed-signal circuits using
  behavioral modeling,'' in \emph{Proc. Int. Conf. Computer-Aided
  Design}.\hskip 1em plus 0.5em minus 0.4em\relax Washington, DC, Nov 1996, pp.
  374--380.

\bibitem{anchor_ANOVA_xiu:2012}
X.~Yang, M.~Choi, G.~Lin, and G.~E. Karniadakis, ``Adaptive {ANOVA}
  decomposition of stochastic incompressible and compressible flows,'' \emph{J.
  Comp. Phys.}, vol. 231, no.~4, pp. 1587--1614, Feb 2012.

\bibitem{HDMR:1999}
H.~Rabitz and O.~F. Alis, ``General foundations of high-dimensional model
  representations,'' \emph{J. Math. Chem.}, vol.~25, no. 2-3, pp. 197--233,
  1999.

\bibitem{anchor_ANOVA_Griebel:2010}
M.~Griebel and M.~Holtz, ``Dimension-wise integration of high-dimensional
  functions with applications to finance,'' \emph{J. Complexity}, vol.~26,
  no.~5, pp. 455--489, Oct 2010.

\bibitem{ANOVA_zqzhang:2012}
Z.~Zhang, M.~Choi, and G.~E. Karniadakis, ``Error estimates for the {ANOVA}
  method with polynomial choas interpolation: tensor product functions,''
  \emph{SIAM J. Sci. Comput.}, vol.~34, no.~2, pp. A1165--A1186, 2012.

\bibitem{anchor_ANOVA_xma:2010}
X.~Ma and N.~Zabaras, ``An adaptive high-dimensional stochastic model
  representation technique for the solution of stochastic partial differential
  equations,'' \emph{J. Compt. Phys.}, vol. 229, no.~10, pp. 3884--3915, May
  2010.

\bibitem{Walter:1982}
W.~Gautschi, ``On generating orthogonal polynomials,'' \emph{SIAM J. Sci. Stat.
  Comput.}, vol.~3, no.~3, pp. 289--317, Sept. 1982.

\bibitem{Ivan:tt_2011}
I.~V. Oseledets, ``Tensor-train decomposition,'' \emph{SIAM J. Sci. Comput.},
  vol.~33, no.~5, pp. 2295--2317, 2011.

\bibitem{Ivan:tt_svd}
I.~V. Oseledets and E.~Tyrtyshnikov, ``Breaking the curse of dimensionality, or
  how to use {SVD} in many dimensions,'' \emph{SIAM J. Sci. Comput.}, vol.~31,
  no.~5, pp. 3744--3759, 2009.

\bibitem{Ivan:tt_across}
------, ``{TT}-cross approximation for multidimensional arrays,'' \emph{Linear
  Alg. Appl.}, vol. 432, no.~1, pp. 70--88, Jan. 2010.

\bibitem{zzhang:JMEMS2014}
Z.~Zhang, M.~Kamon, and L.~Daniel, ``Continuation-based pull-in and lift-off
  simulation algorithms for microelectromechanical devices,'' \emph{J.
  Microelectromech. Syst.}, vol.~23, no.~5, pp. 1084--1093, Oct. 2014.

\bibitem{Golub:1969}
G.~H. Golub and J.~H. Welsch, ``Calculation of {Gauss} quadrature rules,''
  \emph{Math. Comp.}, vol.~23, pp. 221--230, 1969.

\bibitem{cp:Hitchcock}
F.~L. Hitchcock, ``The expression of a tensor or a polyadic as a sum of
  products,'' \emph{J. Math. Phys.}, vol.~6, pp. 39--79, 1927.

\bibitem{cp:Carroll}
J.~D. Carroll and J.~J. Chang, ``Analysis of individual differences in
  multidimensional scaling via an {N}-way generalization of ``{Eckart}-{Young}"
  decomposition,'' \emph{Psychometrika}, vol.~35, pp. 283--319, 1970.

\bibitem{cp:Kiers}
H.~Kiers, ``Towards a standardized notation and terminology in multiway
  analysis,'' \emph{J. Chemometrics}, pp. 105--122, 2000.

\bibitem{tucker:1966}
L.~R. Tucker, ``Some mathematical notes on three-mode factor analysis,''
  \emph{Psychometrika}, vol.~31, no.~5, pp. 279--311, 1966.

\bibitem{tucker:2000}
L.~{De Lathauwer}, B.~{De Moor}, and J.~Vandewalle, ``A multilinear singular
  value decomposition,'' \emph{SIAM J. Matrix Anal.}, vol.~21, pp. 1253--1278,
  2000.

\bibitem{tensor_ill:lim}
V.~{De Silva} and L.-H. Lim, ``Tensor rank and the ill-posedness of the best
  low-rank approximation problem,'' \emph{SIAM J. Sci. Comput.}, vol.~30,
  no.~5, pp. 1084--1127, 2008.

\bibitem{tensor:bigdata}
A.~Cichoki, ``Era of big data processing: A new approach via tensor networks
  and tensor decompositions,'' \emph{arXiv Preprint, arXiv:1403.2048}, March
  2014.

\bibitem{tensor:jmsun}
J.~Sun, D.~Tao, and C.~Faloutsos, ``Beyond streams and graphs: Dynamic tensor
  analysis,'' in \emph{ACM Int. Conf. Knowledge Discovery and Data Mining},
  Aug. 2006, pp. 374--383.

\bibitem{tensor:Kolda2008}
T.~G. Kolda and J.~Sun, ``Scalable tensor decomposition for multi-aspect data
  mining,'' in \emph{IEEE Int. Conf. Data Mining}, 2008, pp. 363--372.

\bibitem{tensor:Vasilescu2002}
M.~A.~O. Vasilescu and D.~Terzopoulos, ``Multilinear analysis of image
  ensembles: Tensorfaces,'' in \emph{Proc. Europ. Conf. Computer Vision}, 2002,
  pp. 447--460.

\bibitem{tensor:suplearning}
D.~Tao, X.~Li, W.~Hu, S.~Maybank, and X.~Wu, ``Supervised tensor learning,'' in
  \emph{Proc. Int. Conf. Data Mining}, 2005, pp. 447--460.

\bibitem{tensor:latentvar}
A.~Anandkumar, R.~Ge, D.~Hsu, S.~M. Kakade, and M.~Telgarsky, ``Tensor
  decompositions for learning latent variable models,'' \emph{arXiv Preprint,
  arXiv:1210.7559}, Oct 2012.

\bibitem{doostan:2009}
A.~Doostan and G.~Iaccarino, ``A least-square approximation of partial
  differential equations with high-dimensional random inputs,'' \emph{J. Comp.
  Physcis}, vol. 228, pp. 4332--4345, 2009.

\bibitem{Nouy:2010}
A.~Nouy, ``Proper generalized decomposition and separated representations for
  the numerical solution of high dimensional stochastic problems,'' \emph{Arch.
  Comp. Meth. Eng.}, vol.~27, no.~4, pp. 403--434, Dec 2010.

\bibitem{Nouy:2009}
A.~Nouy and O.~P. {Le Maitre}, ``Generalized spectral decomposition for
  stochastic nonlinear problems,'' \emph{J. Comp. Phys.}, vol. 228, pp.
  205--235, 2009.

\bibitem{tensor:gelerkin}
B.~N. Khoromskij and C.~Schwab, ``Tensor-structured {Galerkin} approximation of
  parametric and stochastic elliptic {PDEs},'' \emph{SIAM J. Sci. Comput},
  vol.~33, no.~1, pp. 364--385, Oct 2011.

\bibitem{Schwab:2014}
V.~Kazeev, M.~Khammash, M.~Nip, and C.~Schwab, ``Direct solution of the
  chemical master equation using quantized tensor trains,'' \emph{PLOS Comp.
  Biology}, vol.~10, no.~3, pp. e1\,003\,359:1--19, March 2014.

\bibitem{qtt:sc}
B.~N. Khoromskij and I.~Oseledets, ``Quantics-{TT} collocation approximation of
  parameter-dependent and stochastic elliptic {PDEs},'' \emph{Comput. Methods
  in Appl. Math.}, vol.~10, no.~4, pp. 376--394, Jan 2010.

\bibitem{Dolgov:2014}
S.~Dolgov, B.~N. Khoromskij, A.~Litvnenko, and H.~G. Matthies, ``Computation of
  the response surface in the tensor train data format,'' \emph{arXiv preprint,
  arXiv:1406.2816v1}, Jun 2014.

\bibitem{Marzouk:2014}
D.~Bigoni, A.~P. Engsig-Karup, and Y.~M. Marzouk, ``Spectral tensor-train
  decomposition,'' \emph{arXiv preprint, arXiv:1405.5713v1}, May 2014.

\bibitem{ANOVA_sobol:2001}
I.~M. Sobol, ``Global sensitivity indices for nonlinear mathematical models and
  their {Monte} {Carlo} estimates,'' \emph{Math. Comp. Sim.}, vol.~55, no. 1-3,
  pp. 271--280, Feb 2001.

\bibitem{paul:active2013}
P.~G. Constantine, E.~Dow, and Q.~Wang, ``Active subspace methods in theory and
  practice: applications to kriging surfaces,'' \emph{arXiv Preprint,
  arXiv:1304.2070v2}, Dec. 2013.

\bibitem{xli2010}
X.~Li, ``Finding deterministic solution from underdetermined equation:
  large-scale performance modeling of analog/{RF} circuits,'' \emph{IEEE Trans.
  Computer-Aided Design of Integrated Circuits and Systems}, vol.~29, no.~11,
  pp. 1661--1668, Nov 2011.

\bibitem{yxiu:2013}
X.~Yang and G.~E. Karniadakis, ``Reweighted $l_1$ minimization method for
  sothcastic elliptic differential equations,'' \emph{J. Comp. Phys.}, vol.
  248, no.~1, pp. 87--108, Sept. 2013.

\bibitem{JPeng:2014}
J.~Peng, J.~Hampton, and A.~Doostan, ``A weighted $l_1$ minimization approach
  for sparse polynomial chaos expansion,'' \emph{J. Comp. Phys.}, vol. 267,
  no.~1, pp. 92--111, Jun. 2014.

\bibitem{Hampton:2015}
J.~Hampton and A.~Doostan, ``Compressive sampling of sparse polynomial chaos
  expansion: convergence analysis and sampling strategies,'' \emph{J. Comp.
  Phys.}, vol. 280, no.~1, pp. 363--386, Jan. 2015.

\bibitem{ttbox}
I.~V. Oseledets, ``{TT}-{T}oolbox 2.2,'' available online:
  \texttt{http://spring.inm.ras.ru/osel/?page$\_$id=24}.

\bibitem{Dana:2011}
D.~R. Dereus, S.~Natarajan, S.~J. Cunningham, and A.~S. Morris, ``Tunable
  capacitor series/shunt design for integrated tunable wireless front end
  applications,'' in \emph{Proc. IEEE Micro Electro Mechanical Systems}, Jan.
  2011, pp. 805--808.

\bibitem{Stamper:2011}
A.~K. Stamper, C.~V. Jahnes, S.~R. Depuis, A.~Gupta, Z.-X. He, R.~T. Herrin,
  S.~E. Luce, J.~Maling, D.~R. Miga, W.~J. Murphy, E.~J. White, S.~J.
  Cunningham, D.~R. Dereus, I.~Vitomirov, and A.~S. Morris, ``Planar {MEMS}
  {RF} capacitor integration,'' in \emph{Proc. IEEE Solid-State Sensors,
  Actuators Microsyst. Conf.}, Jun. 2011, pp. 1803--1806.

\bibitem{Matt:2013}
M.~Kamon, S.~Maity, D.~DeReus, Z.~Zhang, S.~Cunningham, S.~Kim, J.~McKillop,
  A.~Morris, G.~Lorenz1, and L.~Daniel, ``New simulation and experimental
  methodology for analyzing pull-in and release in {MEMS} switches,'' in
  \emph{Proc. IEEE Solid-State Sensors, Actuators and Microsystems Conference
  (TRANSDUCERS)}, Jun. 2013.

\bibitem{memsp_mannual}
``{MEMS}+ user's mannual,'' {C}oventor, Inc.

\end{thebibliography}

\begin{IEEEbiography}[{\includegraphics[width=1in,height=1.25in,clip,keepaspectratio]{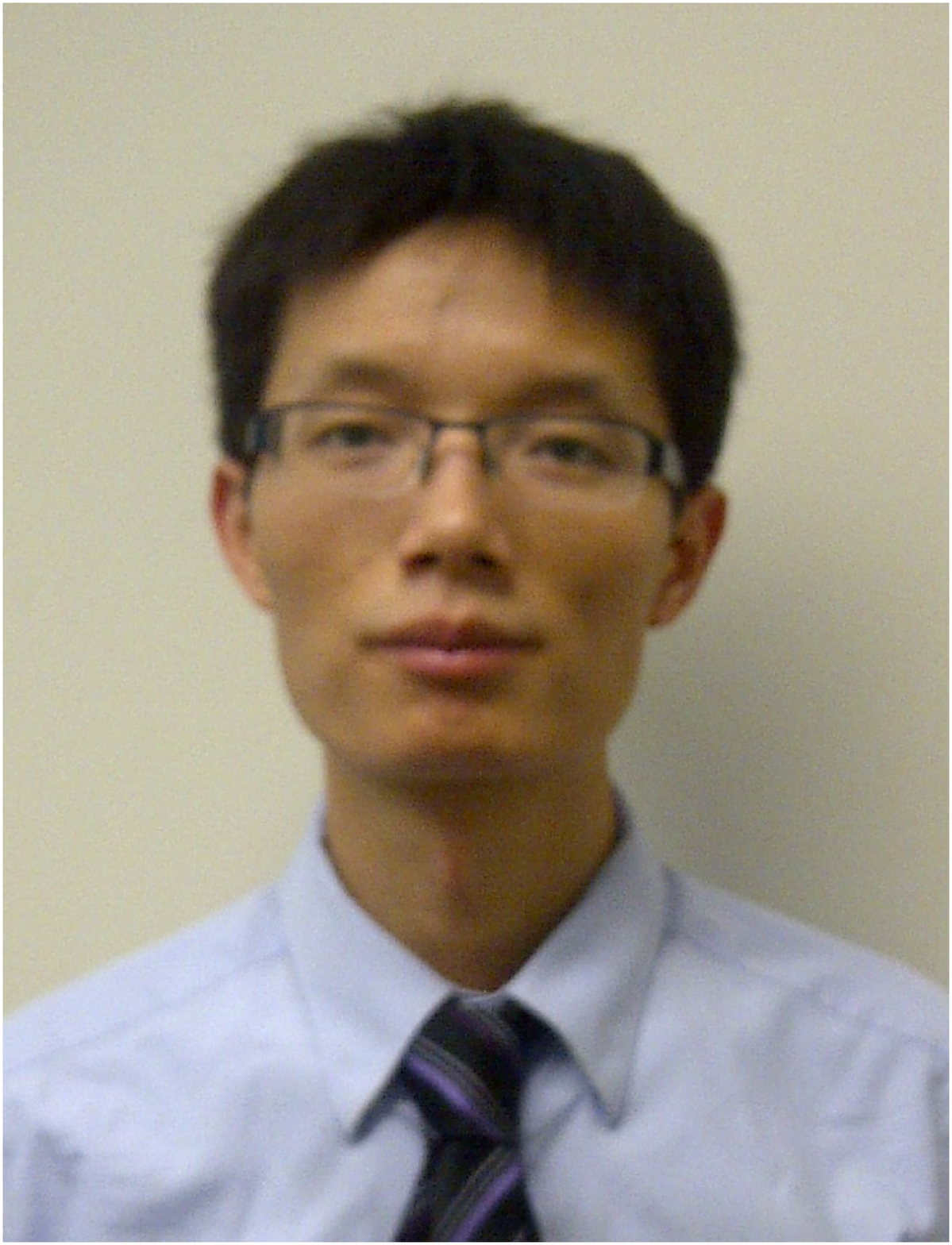}}]{Zheng Zhang} (S'09) received his B.Eng. degree from Huazhong University of Science and Technology, China, in 2008, and M.Phil. degree from the University of Hong Kong, Hong Kong, in 2010. Currently, he is a Ph.D student in Electrical Engineering and Computer Science at the Massachusetts Institute of Technology (MIT), Cambridge, MA. His research interests include uncertainty quantification and tensor analysis, with applications in integrated circuits (ICs), microelectromechanical systems (MEMS), power systems, silicon photonics and other emerging engineering problems.

Mr. Zhang received the 2014 IEEE Transactions on CAD of Integrated Circuits and Systems best paper award, the 2011 Li Ka Shing Prize (university best M.Phil/Ph.D thesis award) from the University of Hong Kong, and the 2010 Mathworks Fellowship from MIT. Since 2011, he has been collaborating with Coventor Inc., working on numerical methods for MEMS simulation. %Some of his research results have been implemented in the commerical MEMS/IC co-design software MEMS+.
\end{IEEEbiography}

\begin{IEEEbiography}[{\includegraphics[width=1in,height=1.25in,clip,keepaspectratio]{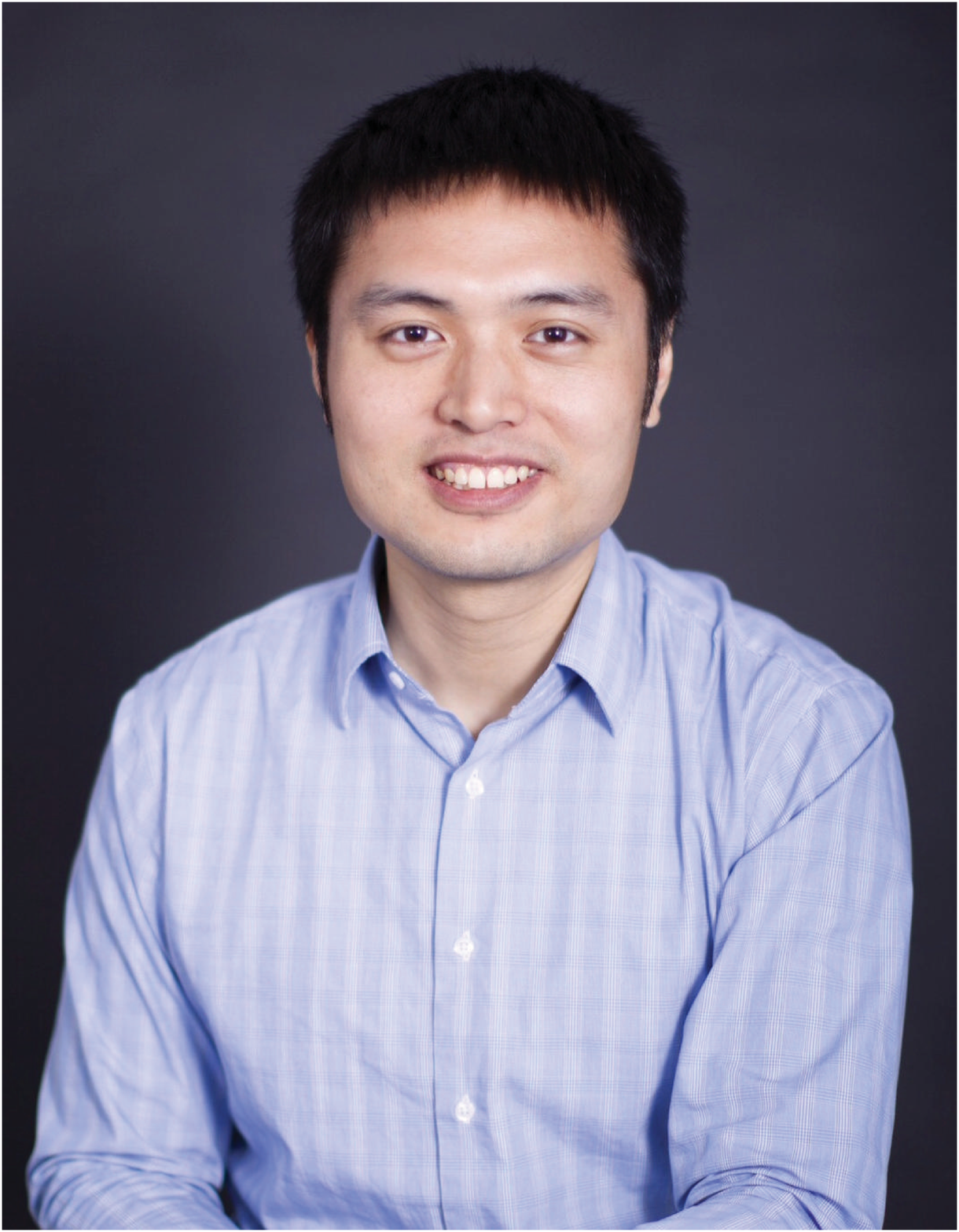}}]{Xiu Yang} received his B.S. and M.S. degrees in applied mathematics from Peking University, China in 2005 and 2008, respectively, and his Ph.D. degree in applied mathematics in 2014 from Brown University, Providence, RI. 

He is a postdoc research associate with the Pacific Northwest National Laboratory, Richland, WA. His research interests include uncertainty quantification, sensitivity analysis, rare events and model calibration with application to multiscale modeling, computational fluid dynamics and electronic engineering. 
\end{IEEEbiography}

\begin{IEEEbiography}[{\includegraphics[width=1in,height=1.25in,clip,keepaspectratio]{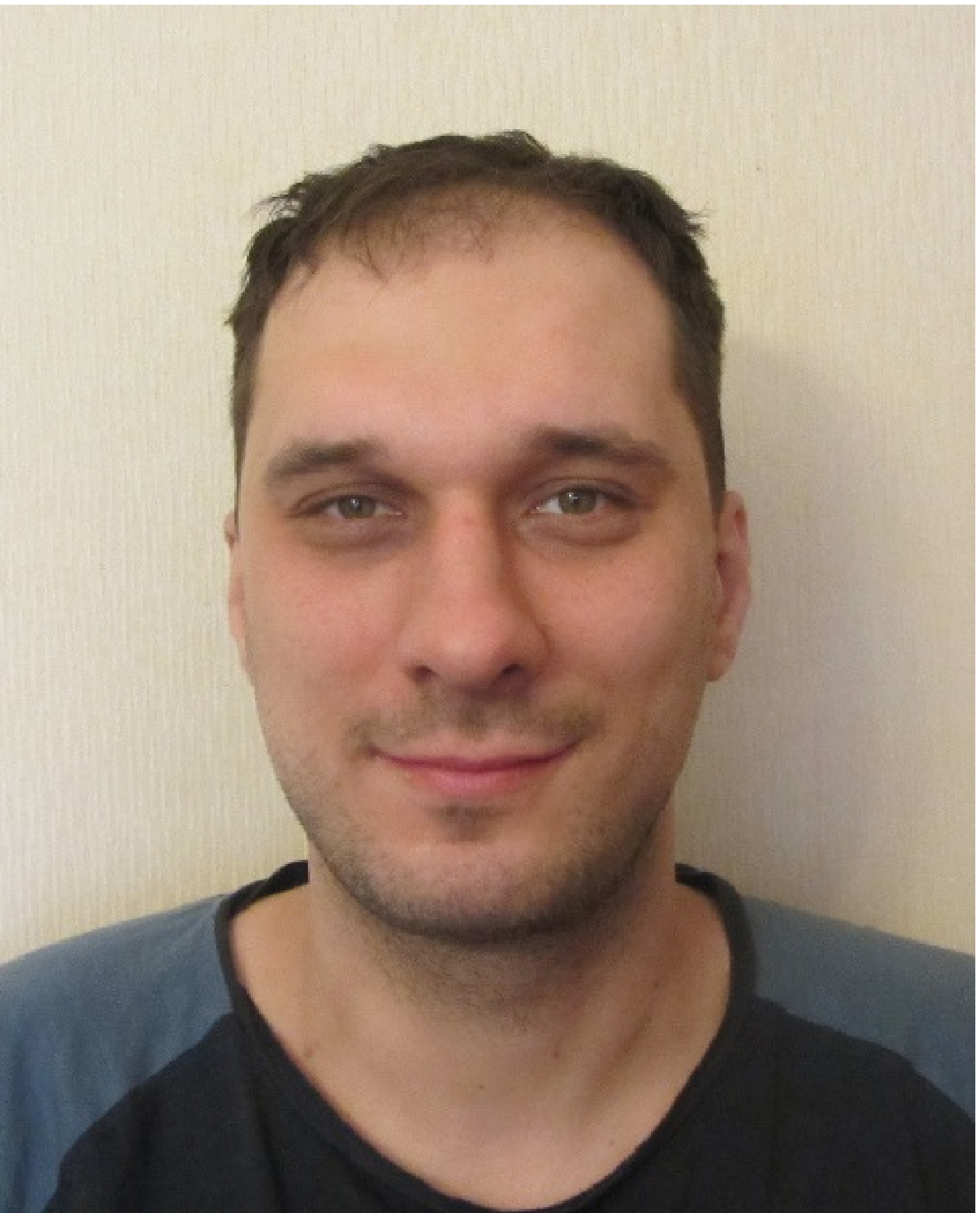}}]{Ivan Oseledets} received his PhD and Doctor of Sciences (second Russian degree) degrees in 2007 and 2012, respectively, both from the Institute of Numerical Mathematics of the Russian Academy of Sciences (INM RAS) in Moscow, Russia.  He has worked in the INM RAS since 2003, where he is now a Leading Researcher.  Since 2013 he has been an Associate Professor in Skolkovo Institute of Science and Technology (Skoltech) in Russia.

His research interests include numerical analysis, linear algebra, tensor methods, high-dimensional problems, quantum chemistry, stochastic PDEs, wavelets, data mining. Applications of interest include solution of integral and differential equations on fine grids, construction of reduced-order models for multi-parametric systems in engineering, uncertainty quantification, \textit{ab initio} computations in quantum chemistry and material design, data mining and compression. 

Dr. Oseledets received the medal of Russian Academy of Sciences for the best student work in Mathematics in 2005; the medal of Russian Academy of Sciences for the best work among young mathematicians in 2009. He is the winner of the Dynasty Foundation contest among young mathematicians in Russia in 2012. 

\end{IEEEbiography}

\begin{IEEEbiography}[{\includegraphics[width=1in,height=1.25in,clip,keepaspectratio]{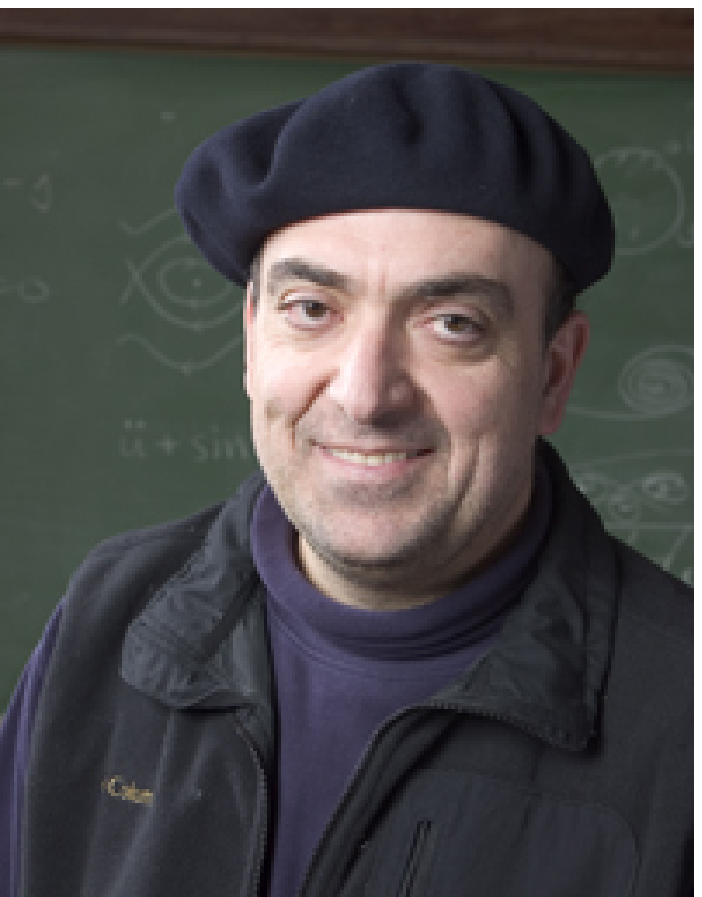}}]{George Karniadakis} received his S.M. (1984) and Ph.D. (1987) from Massachusetts Institute of Technology (MIT). Currently, he is a Full Professor of Applied Mathematics in the Center for Fluid Mechanics at Brown University, Providence, RI. He has been a Visiting Professor and Senior Lecturer of Ocean/Mechanical Engineering at MIT since 2000. His research interests include diverse topics in computational science and engineering, with current focus on stochastic simulation (uncertainty quantification and beyond), fractional PDEs, multiscale modeling of physical and biological systems.  

Prof. Karniadakis is a Fellow of the Society for Industrial and Applied Mathematics (SIAM), Fellow of the American Physical Society (APS), Fellow of the American Society of Mechanical Engineers (ASME) and Associate Fellow of the American Institute of Aeronautics and Astronautics (AIAA). He received the CFD award (2007) and the J Tinsley Oden Medal (2013) by the US Association in Computational Mechanics.  %His H-index is 64 and he has been cited over 20,000 times.
\end{IEEEbiography}

%He is a Fellow of the Society for Industrial and Applied Mathematics (SIAM, 2010-), Fellow of the American Physical Society (APS, 2004-), Fellow of the American Society of Mechanical Engineers (ASME, 2003-) and Associate Fellow of the American Institute of Aeronautics and Astronautics (AIAA, 2006-). He received the CFD award (2007) and the J Tinsley Oden Medal (2013) by the US Association in Computational Mechanics. His h-index is 63 and he has been cited about 20,000 times. 
%
%Karniadakis is the lead Principle Investigator of an OSD/AFOSR MURI on Uncertainty Quantification and Director of a new DOE Center of Mathematics for Mesoscale Modeling of Materials (CM4).} 

\begin{IEEEbiography}[{\includegraphics[width=1in,height=1.25in,clip,keepaspectratio]{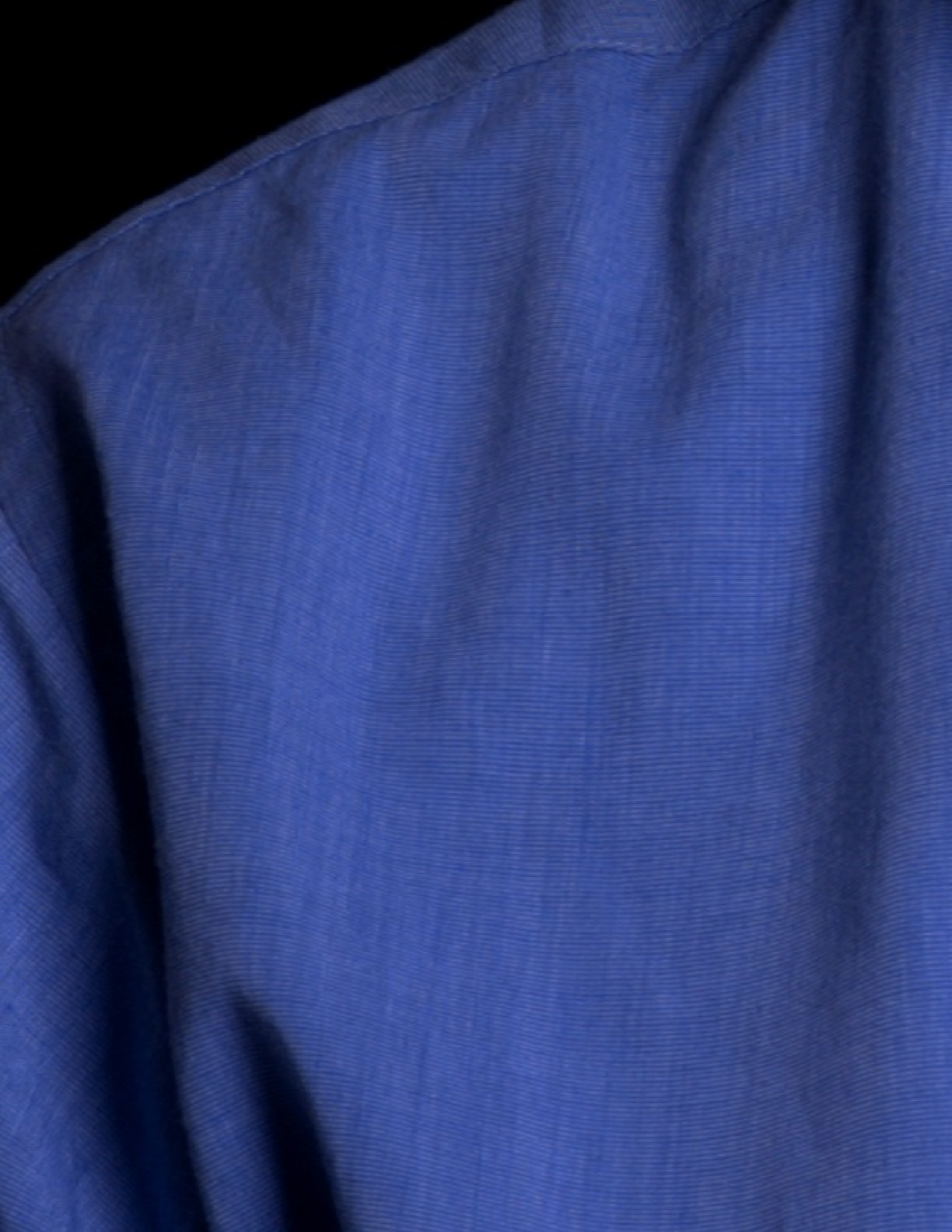}}]{Luca Daniel} (S'98-M'03) received the Ph.D. degree in electrical engineering and computer science from the University of California, Berkeley, in 2003. 

He is currently an Associate Professor in the Electrical Engineering and Computer Science Department of the Massachusetts Institute of Technology (MIT). His research interests include development of integral equation solvers for very large complex systems, uncertainty quantification and stochastic solvers
for large number of uncertainties, and automatic generation of parameterized stable compact models for linear and nonlinear dynamical systems. Applications of interest include simulation, modeling and optimization for mixed-signal/RF/mm-wave circuits, power electronics, MEMs, nanotechnologies, materials, Magnetic Resonance Imaging Scanners, and the human cardiovascular system.    

Prof. Daniel has received the 1999 IEEE Trans. on Power Electronics best paper award; the 2003 best PhD thesis awards from both the Electrical Engineering and the Applied Math departments at UC Berkeley; the 2003 ACM Outstanding Ph.D. Dissertation Award in Electronic Design Automation; 5 best paper awards in international conferences and 9 additional nominations; the 2009 IBM Corporation Faculty Award; the 2010 IEEE Early Career Award in Electronic Design Automation; and the 2014 IEEE Trans. On Computer Aided Design best paper award.
\end{IEEEbiography}

\end{document}